\documentclass[10pt,aps,prd,twocolumn,superscriptaddress,nofootinbib,floatfix]{revtex4-2}

\usepackage{placeins}
\usepackage{amsmath,amssymb}
\usepackage{graphicx}
\usepackage{multirow}
\usepackage{array}
\usepackage{ragged2e}
\usepackage[normalem]{ulem}
\usepackage{siunitx}
\usepackage[colorlinks=true,linkcolor=red,citecolor=green,urlcolor=blue]{hyperref}

\newlength{\colA}
\newlength{\colB}
\newlength{\colC}

\begin{document}

\title{Fermion Mass Hierarchies and the Exceptional Jordan Algebra}

\author{Bishnu Gupta Teli}\email{ph24c802@smail.iitm.ac.in}
\affiliation{Department of Physics, Indian Institute of Technology Madras, Chennai 600036, India}

\author{Tejinder P. Singh}\email{tpsingh@tifr.res.in}
\affiliation{Tata Institute of Fundamental Research, Homi Bhabha Road, Mumbai 400005, India}

\date{\today}

\begin{abstract}
We develop a spectral framework for fermion mass hierarchies based on the exceptional Jordan algebra $J_3(\mathbb{O}_{\mathbb{C}})$. Starting from an octonionic state-space construction in $\mathbb{C}\otimes\mathbb{O}$, we embed the selected family-labelled states into Hermitian Jordan elements whose eigenvalues define intrinsic spectral invariants. The ordered spectral scales generate cubic ladder structures in the symmetric representation $\mathrm{Sym}^3(\mathbf{3})$, and the consistency of multiplicative-hierarchy composition naturally yields power-law relations between fermion masses and spectral scales. The construction should be viewed as a phenomenological spectral deformation of the rigid exceptional-Jordan framework discussed below: we retain selected cubic-ladder assignments and a prescribed cross-sector mapping as phenomenological inputs, but promote the relative normalization, hierarchy exponent, and charged-lepton octonionic phase to fitted spectral moduli. A global logarithmic fit to six charged-fermion mass ratios at $\mu=M_Z$ lowers the unpenalized log-residual relative to the rigid point, mainly through the top-to-charm ratio, while the individual ratios are not uniformly improved. The best-fit hierarchy exponent remains close to the square-root scaling regime, $p\simeq1$. In the neutrino sector, the framework accommodates both normal and inverted ordering while remaining consistent with oscillation data and current cosmological bounds on the total neutrino mass. Thus, the proposal is a phenomenological spectral parametrization of fermion mass hierarchies based on invariants of the exceptional Jordan algebra, rather than a first-principles derivation of Standard Model fermion masses.
\end{abstract}

\maketitle

\section{Introduction}
\label{sec:intro}
Understanding the origin of fermion masses and their hierarchical structure remains one of the central open problems in particle physics. While the Standard Model successfully parametrizes fermion masses through Yukawa couplings, it does not explain the observed pattern of mass ratios across generations. These ratios span several orders of magnitude and exhibit nontrivial structure, suggesting an underlying organizing principle beyond the Standard Model. This has led to a long-standing line of research exploring whether the structure of fundamental particles is governed by underlying algebraic frameworks, particularly normed division algebras. The natural appearance of $\mathbb{C}$ in quantum mechanics and $\mathbb{H}$ in spin and gauge structures raises the question of whether the largest division algebra, the octonions $\mathbb{R} \subset \mathbb{C} \subset \mathbb{H} \subset \mathbb{O}$, also plays a role in particle physics, a connection that can be traced back to the work of Günaydin and Gürsey~\cite{Gunaydin:1973rs, Baez:2001dm}, who showed that octonions provide a natural framework for quark color symmetry. Subsequent developments by Dixon~\cite{dixon2013division} demonstrated that tensor products such as $\mathbb{C} \otimes \mathbb{H} \otimes \mathbb{O}$ encode many structural aspects of the Standard Model, including gauge symmetries and fermion representations. More recently, Furey~\cite{Furey:2010fm, Furey:2015yxg} has shown that one generation of fermions can be identified with minimal left ideals of $\mathbb{C} \otimes \mathbb{O}$, while Quinta~\cite{Quinta:2025kal} derived the three-family structure from triality in $C\ell_{8,0}$. Earlier connections between triality, exceptional groups, and fermion generations were also discussed by Ramond~\cite{Ramond:1977aw}. Further developments have explored how the exceptional Jordan algebra and related octonionic geometric frameworks may encode these structural features~\cite{Boyle:2020ctr, Springer2000, toupin2026division, Gresnigt:2026eju, Gresnigt:2026saa}.

Despite these structural successes, a generally accepted mechanism mapping algebraic invariants to quantitative fermion mass ratios remains missing. Earlier Jordan-algebraic approaches~\cite{Bhatt:2021cpg, Singh:2025xxv} proposed candidate relations for organizing fermion mass hierarchies. In particular, Ref.~\cite{Singh:2025xxv} presents a conditional mass-ratio ansatz based on a symmetric Jordan spectrum and selected ladders in $\mathrm{Sym}^3(\mathbf{3})$. In that ansatz, the choice $\delta^2=3/8$ specifies a normalization; it is not derived as a physical mass scale from exceptional symmetry alone. The identification of particular weight monomials with physical fermion masses is an additional phenomenological assumption. Likewise, the relation between the down-quark and charged-lepton sectors requires a cross-sector prescription. It does not follow from a Dynkin-diagram automorphism alone. The present work studies a fitted phenomenological deformation of this conditional construction. We retain selected cubic-ladder assignments and allow the relative normalization $r$, hierarchy exponent $p$, and charged-lepton phase $\Phi_e$ to vary in a global fit to charged-fermion mass-ratio data. The purpose is to investigate the effect of these deformations, not to turn the underlying conditional mass relations into a first-principles derivation of the Standard Model fermion masses.

Starting from an octonionic state-space construction in $\mathbb{C}\otimes\mathbb{O}$, we embed selected family-labelled states into Hermitian elements of $J_3(\mathbb{O}_{\mathbb{C}})$, using the family assignment as a phenomenological ansatz. The resulting Hermitian matrices define intrinsic spectral invariants, whose eigenvalues provide three ordered scales $(a,b,c)$ in the real-invariant sectors used below, or ordered positive magnitudes in the fully complexified setting. We show that fermion mass ratios can be constructed from these spectral scales through a cubic ladder structure in the symmetric representation $\mathrm{Sym}^3(\mathbf{3})$. Assuming that mass contrasts depend only on spectral ratios and satisfy multiplicative composition, the mass-to-scale relation takes a power-law form. Different fermion sectors are assigned to selected ladders in the cubic weight diagram. The charged-lepton ladder is related to the down-quark ladder by a prescribed $b\leftrightarrow c$ weight-diagram reflection together with an additional cross-sector transport assumption. The best-fit hierarchy exponent lies close to the square-root scaling limit $p\simeq1$. In the neutrino sector, the construction accommodates both normal and inverted ordering while remaining consistent with oscillation data and cosmological bounds on the total neutrino mass.

It is important to emphasize that the present construction is not derived from a complete dynamical theory, but instead provides an effective spectral organization of fermion mass hierarchies in terms of exceptional Jordan invariants. The octonionic Clifford construction organizes candidate state labels and occupation-number gradings, while the spectral geometry of Hermitian Jordan elements provides the quantities used to parametrize fermion mass hierarchies. The identification of these structures with the complete Standard Model gauge representations is not derived in the present work. Further, the octonionic Clifford construction is not used as a conventional grand-unified framework, but rather as an algebraic starting point for organizing flavor structure. The exceptional Jordan algebra $J_3(\mathbb O_{\mathbb C})$ is employed as a spectral arena in which inter-generation structure is encoded through Hermitian Jordan elements and their intrinsic spectral invariants. The resulting construction should therefore be interpreted as a framework for fermion mass hierarchies rather than as a standard $E_6$ unification scheme. Throughout, we distinguish algebraic statements that follow from the Jordan and octonionic structures from phenomenological assumptions that identify spectral quantities with physical fermion masses.

A concise comparison with Ref.~\cite{Singh:2025xxv} is given in Table~\ref{tab:comparison-ref9}. The cited work formulates conditional mass-ratio relations under specified spectral, weight-assignment, and cross-sector assumptions. The present paper studies a phenomenological deformation of selected parts of this construction by fitting the relative normalization, hierarchy exponent, and charged-lepton phase. The resulting fit is assessed as a spectral parametrization rather than as a first-principles derivation of the physical fermion mass spectrum. The present analysis also considers the compatibility of the spectral description with the two standard neutrino mass orderings.

\begin{table*}[htbp]
\centering

\setlength{\tabcolsep}{4pt}
\renewcommand{\arraystretch}{1.5}
\setlength{\colA}{0.14\textwidth}
\setlength{\colB}{0.4\textwidth}
\setlength{\colC}{0.38\textwidth}

\begin{tabular}{lll}
\hline\hline
\begin{minipage}[t]{\colA}\textbf{Aspect}\end{minipage} &
\begin{minipage}[t]{\colB}\textbf{Ref.~\cite{Singh:2025xxv}}\end{minipage} &
\begin{minipage}[t]{\colC}\textbf{Present work}\end{minipage}
\\[0.4em]
\hline
\begin{minipage}[t]{\colA}Status\end{minipage} &
\begin{minipage}[t]{\colB}\justifying\noindent Conditional mass-ratio ansatz; closed-form relations under specified spectral and weight-assignment assumptions. \end{minipage} &
\begin{minipage}[t]{\colC}\justifying\noindent Effective spectral deformation of the rigid point; $r$, $p$, and $\Phi_e$ are fitted to six charged-fermion ratios.\end{minipage}
\\[0.8em]
\begin{minipage}[t]{\colA}Shared structure\end{minipage} &
\begin{minipage}[t]{\colB}\justifying\noindent Selected cubic $\mathrm{Sym}^3(\mathbf3)$ ladders and a prescribed cross-sector mapping between quarks and leptons. \end{minipage} &
\begin{minipage}[t]{\colC}\justifying\noindent Selected ladder assignments and a prescribed cross-sector mapping are retained as phenomenological assumptions. \end{minipage}
\\[0.8em]
\begin{minipage}[t]{\colA}Spectral data\end{minipage} &
\begin{minipage}[t]{\colB}\justifying\noindent Symmetric spectrum $(s-\delta,s,s+\delta)$ with $\delta^2=3/8$ imposed as a normalization choice. \end{minipage} &
\begin{minipage}[t]{\colC}\justifying\noindent Jordan spectral invariants with relative normalization $r$ and nonzero octonionic triple-product phase through $\tau=\Re(y(xz))$.\end{minipage}
\\[0.8em]
\begin{minipage}[t]{\colA}Phenomenology\end{minipage} &
\begin{minipage}[t]{\colB}\justifying\noindent Conditional charged-fermion mass-ratio relations under the
stated spectral and cross-sector assumptions. \end{minipage} &
\begin{minipage}[t]{\colC}\justifying\noindent Global logarithmic fit of charged-fermion ratios at $\mu=M_Z$; CKM and PMNS mixing are outside the scope of this paper.\end{minipage}
\\[0.8em]
\begin{minipage}[t]{\colA}Interpretation\end{minipage} &
\begin{minipage}[t]{\colB}\justifying\noindent Conditional benchmark relations under specified assumptions. \end{minipage} &
\begin{minipage}[t]{\colC}\justifying\noindent Nearby phenomenological moduli-space probe; the unpenalized global log-residual is lower, but individual ratios are mixed, and the parameter count must be kept explicit.\end{minipage}
\\[0.4em]
\hline\hline
\end{tabular}
\caption{Structural comparison between the rigid framework of Ref.~\cite{Singh:2025xxv} and the present spectral deformation.}
\label{tab:comparison-ref9}
\end{table*}

The remainder of this paper is organized as follows. In Sec.~\ref{sec:fermion_generations}, we review the octonionic construction of fermion generations and the associated Clifford-algebraic structure. In Sec.~\ref{sec:hermitian_embedding}, the three-generation system is embedded into Hermitian elements of the exceptional Jordan algebra $J_3(\mathbb O_{\mathbb C})$. Sec.~\ref{sec:constrained-embeddings} analyzes the constrained Jordan sector and the residual moduli structure of the embedding. In Sec.~\ref{sec:spectral_construction}, we derive the spectral hierarchy relations from cubic ladder structures in $\mathrm{Sym}^3(\mathbf 3)$. Sec.~\ref{sec:results} presents the resulting fermion mass hierarchies and their phenomenological comparison with experimental data, including the neutrino sector. Finally, Sec.~\ref{sec:conclusion} discusses the interpretation, limitations, and possible dynamical implications of the framework.

\section{Octonionic Eigenstructure and Fermion Generations}
\label{sec:fermion_generations}
The guiding principle of this construction is that internal particle representations arise as generalized ideals of an algebra acting on itself, as proposed in Ref.~\cite{Furey:2010fm}. Specifically, we use an eigenspace decomposition of $\mathbb{C}\otimes\mathbb{O}$ to organize an eight-dimensional complex state space with proposed one-generation fermion labels. Identifying this state space with a complete Standard Model generation requires consistent representation and charge assignments beyond the eigenspace decomposition alone. Because the octonions are non-associative, as discussed in Appendix~\ref{app:octonions}, sequences of multiplications must be treated as linear operators acting on $\mathbb{C}\otimes\mathbb{O}$. Following Furey~\cite{Furey:2010fm, Furey:2015yxg}, we denote left multiplication by an octonionic unit $e_i$ using an arrow notation:
\begin{equation}
\overleftarrow{e_i}\,v \equiv e_i v ,
\end{equation}
and more generally chains such as $\overleftarrow{e_i e_j}$ denote successive left multiplications $v\mapsto e_i(e_j v)$. Using this notation, we consider the operator
\begin{equation}
    \mathcal O v = c\, v, \qquad \mathcal O = \frac{i \overleftarrow{e_7}}{2}.
\end{equation}
Since $e_7^2=-1$, we have $\mathcal O^2=\tfrac14$, hence
\begin{equation}
    c=\pm\tfrac12 .
\end{equation}
The algebra decomposes into eigenspaces
\begin{equation}
    \mathbb{C}\otimes\mathbb{O}=\mathcal H_{+\frac12}\oplus\mathcal H_{-\frac12},
\end{equation}
each four-dimensional over $\mathbb{C}$. An explicit eigenbasis is
\begin{gather}
    \nu = \frac{1+ i e_7}{2}, \quad e^- = \frac{1- i e_7}{2}, \\
    u_r = \frac{-e_5+ie_4}{2}, \quad u_g = \frac{-e_3+ie_1 }{2}, \quad u_b = \frac{-e_6+ie_2}{2},\\
    d_r = \frac{e_5+ie_4}{2}, \quad d_g = \frac{e_3+ie_1 }{2}, \quad d_b = \frac{e_6+ie_2}{2}.
\end{gather}
These eight states are used as a proposed one-generation state basis in the present construction. Their identification with a complete Standard Model generation requires consistent particle, colour, weak-isospin, and charge assignments. Following Refs.~\cite{Furey:2010fm, Furey:2015yxg}, antiparticle states in this construction are obtained by complex conjugation, $i \mapsto -i$, which leaves the octonionic basis elements unchanged and maps particle states to their corresponding antiparticles in $\mathbb{C} \otimes \mathbb{O}$. The relevant octonionic symmetry is the automorphism group $G_2=\mathrm{Aut}(\mathbb{O})$, which acts on $\mathbb{O}$. Fixing $e_7$ selects the stabilizer subgroup
\begin{equation}
    \mathrm{Stab}_{G_2}(e_7)\cong SU(3).
\end{equation}
Under this action,
\begin{equation}
    \mathbb{C}\otimes\mathbb{O} = \mathbf{1}\oplus\mathbf{1}\oplus\mathbf{3}\oplus\bar{\mathbf{3}}.
    \label{eq:colour_decomposition}
\end{equation}
Here $\mathrm{span}_{\mathbb{C}}\{1,e_7\}\cong \mathbf{1}\oplus\mathbf{1}$ is the two-dimensional singlet sector, and the remaining six states transform as $\mathbf{3}\oplus\bar{\mathbf{3}}$. The operator
\begin{equation}
    \mathcal O=\frac{i\overleftarrow{e_7}}{2}
\end{equation}
acts diagonally on the corrected state basis. Its eigenvalues are $+\frac12$ on $\{\nu,u_r,u_g,u_b\}$ and $-\frac12$ on $\{e^-,d_r,d_g,d_b\}$. We retain $\mathcal O$ as an auxiliary grading operator, rather than identifying it with the physical weak-isospin generator $T_3$. The eigenspace splitting alone does not establish the physical weak-doublet action on the original eight-state module. A conditional weak-isospin action on an enlarged matter space is specified below.

Independently, the octonionic left-multiplication construction provides fermionic ladder operators and an associated occupation-number grading. Following Furey’s construction~\cite{Furey:2010fm, Furey:2015yxg}, one introduces ladder operators inside $\mathbb{C}\otimes\mathbb{O}$ generating a representation of the Clifford algebra $\mathbb{C}\ell(6)$. Choose a fixed complex direction $e_a$ and define
\begin{equation}
    \alpha_1=\frac{1}{2}(\overleftarrow{e_{b_1}}+i \overleftarrow{e_{c_1}}), \quad
    \alpha_2=\frac{1}{2}(\overleftarrow{e_{b_2}}+i \overleftarrow{e_{c_2}}), \quad
    \alpha_3=\frac{1}{2}(\overleftarrow{e_{b_3}}+i \overleftarrow{e_{c_3}}),
\end{equation}
where $(e_{b_k},e_{c_k},e_a)$ form the oriented quaternionic triples in the Fano plane. These operators satisfy fermionic anticommutation relations
\begin{equation}
    \{\alpha_i,\alpha_j\}=0, \qquad \{\alpha_i,\alpha_j^\dagger\}=\delta_{ij},
\end{equation}
and hence generate $\mathbb{C}\ell(6)$. The ladder operators generate the Clifford-algebraic occupation-number structure used in related constructions~\cite{Furey:2010fm, Furey:2015yxg}. Identifying the particular eight-state module used here with the complete chiral Standard Model matter content requires additional representation and gauge assignments, as specified below. The associated number operator is
\begin{equation}
    N = \sum_{i=1}^{3} \alpha_i^\dagger \alpha_i,
\end{equation}
whose eigenvalues are $\{0,1,2,3\}$, corresponding to occupation numbers of the states. We define the normalized occupation-number grading
\begin{equation}
    Q_{\mathrm{occ}}=\frac{N}{3}.
\end{equation}
Its eigenvalues are
\begin{equation}
    Q_{\mathrm{occ}}=0,\ \frac{1}{3},\ \frac{2}{3},\ 1.
\end{equation}
A related number-operator construction, in which $N/3$ mirrors the electric-charge pattern, was studied by Furey~\cite{Furey:2015tqa}. In the present construction, however, these eigenvalues are retained as an occupation-number grading; identifying this grading with physical electric charge requires a consistent assignment of the particle and antiparticle states and their gauge quantum numbers. For a Standard Model embedding, the physical electric-charge operator $Q_{\mathrm{em}}$, the weak-isospin generator $T_3$, and the hypercharge generator $Y$ must satisfy
\begin{equation}
    Q_{\mathrm{em}}=T_3+\frac{Y}{2}.
    \label{eq:ew}
\end{equation}
For the original eight-state module, this relation is a compatibility requirement rather than a derived identity: the physical weak-isospin generator $T_3$ and hypercharge generator $Y$ are not constructed on that module. The conditional enlarged benchmark below specifies them as additional inputs. The results established above are the $SU(3)$ stabilizer associated with the chosen octonionic direction, the fermionic Clifford-algebra construction with its occupation-number grading, and the $\mathcal O$ eigenspace splitting of the corrected state basis. These results do not, by themselves, establish a representation of the full product group
\begin{equation}
    SU(3)_c\times SU(2)_L\times U(1)_Y.
\end{equation}
The physical weak-isospin action and hypercharge assignments are not derived on the original eight-state module. A conditional enlarged matter benchmark, with these assignments specified as additional inputs, is given below.

For the fixed colour action in Eq.~\eqref{eq:colour_decomposition}, the original eight-state module contains a triplet and an anti-triplet. Since
\begin{equation}
    \operatorname{Hom}_{SU(3)} (\bar{\mathbf{3}},\mathbf{3})=\{0\},
\end{equation}
a colour-equivariant weak generator cannot pair these inequivalent sectors into the required quark weak doublets. Related division-algebraic ladder-operator constructions of the broader Standard Model gauge symmetry have been studied by Furey~\cite{Furey:2018drh}. We therefore introduce an enlarged matter assignment as a conditional physical benchmark, not as a consequence of the original $\mathbb{C}\otimes\mathbb{O}$ construction. The occupation-number grading is motivated by the number-operator construction in Refs.~\cite{Furey:2010fm, Furey:2015yxg}. For the conditional benchmark, take two independent eight-dimensional Clifford coefficient modules $\mathcal I_+$ and $\mathcal I_-$, with conjugate colour and electric-charge assignments. Their occupation degrees $n=0,1,2,3$ have dimensions $1,3,3,1$:
\begin{align}
\mathcal I_+&=(\mathbf{1},0)\oplus(\bar{\mathbf{3}},\tfrac13)\oplus(\mathbf{3},\tfrac23)\oplus(\mathbf{1},1),\\
\mathcal I_-=&(\mathbf{1},0)\oplus(\mathbf{3},-\tfrac13)\oplus(\bar{\mathbf{3}},-\tfrac23)\oplus(\mathbf{1},-1).
\label{eq:conditional_modules}
\end{align}
Each pair denotes the colour representation and electric charge. We take $Q_{\mathrm{em}}=N/3$ on $\mathcal I_+$ and $Q_{\mathrm{em}}=-N/3$ on $\mathcal I_-$. The two modules are independent; their combination and the weak pairing introduced below are additional assumptions.
The quark doublet pairs the charge-$+\frac23$ triplet in $\mathcal I_+^{(2)}$ with the charge-$-\frac13$ triplet in $\mathcal I_-^{(1)}$. The lepton doublet pairs the neutral singlet in $\mathcal I_+^{(0)}$ with the charge-$-1$ singlet in $\mathcal I_-^{(3)}$. On these selected doublets, define
\begin{align}
T_+ =&|\nu_L\rangle\langle e_L|+\sum_{a=r,g,b}|u_L^a\rangle\langle d_L^a|,\qquad T_- = T_+^\dagger,\\
T_3&=\frac12[T_+,T_-],\qquad Y=2(Q_{\mathrm{em}}-T_3).
\label{eq:conditional_weak}
\end{align}
These are new block operators on the enlarged space, not the original octonionic grading. The remaining states are weak singlets. The resulting left-handed multiplets have the following quantum numbers:
\begin{center}
\begin{tabular}{lcr}
\hline\hline
Field & $(SU(3)_c,SU(2)_L)$ & $Y$\\
\hline
$Q_L=(u_L,d_L)$ & $(\mathbf{3},\mathbf{2})$ & $\frac13$\\
$L_L=(\nu_L,e_L)$ & $(\mathbf{1},\mathbf{2})$ & $-1$\\
$u_L^c$ & $(\bar{\mathbf{3}},\mathbf{1})$ & $-\frac43$\\
$d_L^c$ & $(\bar{\mathbf{3}},\mathbf{1})$ & $\frac23$\\
$e_L^c$ & $(\mathbf{1},\mathbf{1})$ & $2$\\
$\nu_L^c$ (optional) & $(\mathbf{1},\mathbf{1})$ & $0$\\
\hline\hline
\end{tabular}
\end{center}
Here, the superscript $c$ denotes the left-handed charge conjugate of a right-handed physical field. These assignments satisfy $Q_{\mathrm{em}}=T_3+Y/2$ on the weak doublets and $Q_{\mathrm{em}}=Y/2$ on the weak singlets. For a Higgs doublet with $Y_H=1$, the hypercharge sums for the usual Yukawa operators in all-left-handed notation vanish:
\begin{align}
Y(Q_L)+Y_H+Y(u_L^c)&=\frac13+1-\frac43=0,\\
Y(Q_L)-Y_H+Y(d_L^c)&=\frac13-1+\frac23=0,\\
Y(L_L)-Y_H+Y(e_L^c)&=-1-1+2=0.
\end{align}
If $\nu_L^c$ is included, the Dirac-neutrino condition is $Y(L_L)+Y_H+Y(\nu_L^c)=-1+1+0=0$. The weak pairing, chirality, and independent module copies are assumptions. For three families, an additional family multiplicity is required; its selection is not derived here. For the mass-ratio analysis below, we assume this chiral matter content as the target physical field content. The Jordan spectral ansatz is used phenomenologically to parametrize family-dependent coefficients of gauge-invariant Yukawa operators. The map from the spectral quantities to canonically normalized Yukawa singular values remains to be established. This benchmark does not determine the spectral centres, the relative normalization $r$, or the mass functional.

The family-label ansatz considered here uses inequivalent complex structures within the octonions. A choice of imaginary unit $e_a$ defines a complex structure on $\mathbb{C}\otimes\mathbb{O}$. Distinct choices related by automorphisms of $\mathbb{O}$ correspond to equivalent algebraic constructions but inequivalent complex structures. Among these, sets of mutually non‑collinear imaginary units forming cyclic triples in the Fano plane play a distinguished role. In the present construction, the selected non-collinear triple is used as an ansatz for assigning three family labels. We do not establish a unique minimal $G_2$ orbit or derive the physical number of fermion generations from the octonionic algebra alone. The explicit cyclic non-collinear triple $\{e_6,e_3,e_5\}$ is the selected family-label ansatz used below.\footnote{Here ``non-collinear'' means that the three selected imaginary directions do not lie on a single line of the Fano plane, so that no product of two selected directions is proportional to the third. The choice of such a triple is not unique; any other admissible non-collinear choice gives an equivalent realization of the same octonionic construction, related by an automorphism of $\mathbb O$.} Its members define the three complex structures considered in the spectral embedding; their physical interpretation as distinct generations is an additional assumption. For this triple, define for each $a\in\{6,3,5\}$
\begin{equation}
    \mathcal{O}_a = \frac{i\overleftarrow{e_a}}{2}, \qquad \mathcal{O}_a^2=\tfrac14 .
\end{equation}
Each $\mathcal{O}_a$ defines a decomposition
\begin{equation}
    \mathbb{C}\otimes\mathbb{O} = \mathcal H^{(a)}_{+\frac12} \oplus \mathcal H^{(a)}_{-\frac12}.
\end{equation}
The associated eigenstates are
\begin{align}
    \nu^{(a)} &= \frac{1+ i e_a}{2}, &
    e^{(a)} &= \frac{1- i e_a}{2}, \\
    u^{(a)}_k &= \frac{-e_{c_k} + i e_{b_k}}{2}, & d^{(a)}_k &= \frac{e_{c_k} + i e_{b_k}}{2}.
\end{align}
where $(e_{b_k},e_{c_k},e_a)$ form the oriented quaternionic triples in the Fano plane. For each $a$, this yields a labelled eight-state basis used in the family ansatz
\begin{equation}
    (\nu^{(a)},e^{(a)};u^{(a)}_k,d^{(a)}_k)\qquad k\in\{r,g,b\}.
\end{equation}
If $g\in G_2$ maps $e_a\mapsto e_b$,
\begin{equation}
    \mathcal{O}_b = g \mathcal{O}_a g^{-1}.
\end{equation}
The three decompositions are related by $G_2$ automorphisms. In the present construction, the selected triple is used as an ansatz for assigning three family labels; the choice does not by itself derive the number of physical generations. The cyclic Fano-plane triple \(\{e_6,e_3,e_5\}\) therefore provides a natural three-family ansatz within the octonionic construction. We do not claim that \(G_2\) dynamically selects exactly three physical generations; rather, the chosen cyclic triple supplies the family labels used in the spectral embedding below. Related Spin(8)-triality considerations appear in other algebraic approaches to flavor structure~\cite{toupin2026division}.

\section{Hermitian Embedding of Fermion Families}
\label{sec:hermitian_embedding}
For each complex structure $a\in\{6,3,5\}$ the states
\begin{equation}
    (\nu^{(a)}, e^{(a)}; u^{(a)}_k, d^{(a)}_k),\qquad k\in\{r,g,b\},
\end{equation}
form a basis of $\mathbb{C}\otimes\mathbb{O}$. The algebraic construction fixes only relative charge eigenvalues; the overall normalization remains free. Since color indices transform independently of the family structure, we suppress color by fixing a single representative, chosen to be $k=r$. All subsequent spectral constructions are therefore performed in the reduced basis
\begin{equation}
    (\nu^{(a)}, e^{(a)}; u^{(a)}, d^{(a)}),
\end{equation}
with color understood. We introduce a family-wise normalization
\begin{equation}
    \psi_f^{(a)} \;\longrightarrow\;\frac{e^{i\alpha_{a,f}}}{2}\,\psi_f^{(a)}, \qquad \alpha_{a,f}\in\mathbb{R},
\end{equation}
such that each fermion in a given family contributes equally to the total norm while the overall normalization satisfies $\sum_f \|\psi_f^{(a)}\|^2 = 1$.\footnote{Here, $\|x\|^2 \equiv \operatorname{Sc}(x x^\dagger)$ denotes the algebraic quadratic form on the complexified octonions $\mathbb{O}_\mathbb{C}$ (see Appendix~\ref{app:J3OC}).} The constant denominator is common to all states, while the phases $\alpha_{a,f}$ may depend on both the generation index $a$ and the fermion species $f$. This fixes the overall scale while preserving relative phases between different states.

To obtain intrinsic spectral invariants, the three-generation space is embedded into the exceptional Jordan algebra $J_3(\mathbb{O}_{\mathbb{C}})$. The connection between octonionic projective geometry and $J_3(\mathbb{O})$ is well established~\cite{Baez:2001dm, Springer2000, Todorov:2018yvi}. Exceptional structures of this type have long been considered in attempts to organize fermionic multiplets within unified algebraic frameworks~\cite{Ramond:1977aw}. For each fermion type $f\in\{\nu,e,u,d\}$ we define
\begin{equation}
    A_f =
    \begin{pmatrix}
        s_f & x_f & y_f^\dagger \\
        x_f^\dagger & s_f & z_f \\
        y_f & z_f^\dagger & s_f
    \end{pmatrix},
    \qquad
    s_f\in\mathbb{C},\; x_f,y_f,z_f\in\mathbb{O}_{\mathbb{C}},
\end{equation}
where the off-diagonal entries encode the three generations:
\begin{equation}
    (x_f, y_f, z_f) = \big(\psi_f^{(6)}, \psi_f^{(3)}, \psi_f^{(5)}\big).
    \label{eq:states}
\end{equation}
The diagonal term is family-invariant, while the off-diagonal entries encode inter-generation structure. Each matrix decomposes into
\begin{equation}
    A_f = D_f + O_f ,
\end{equation}
where
\begin{equation}
    D_f = s_f\,\mathbf{1}_3 ,
    \qquad
    O_f =
    \begin{pmatrix}
        0 & x_f & y_f^\dagger \\
        x_f^\dagger & 0 & z_f \\
        y_f & z_f^\dagger & 0
    \end{pmatrix}.
\end{equation}
The off-diagonal component $O_f$ encodes the inter-generation structure determined by the three inequivalent complex directions of the octonions. Since the electric charge operator is defined only up to overall normalization, the relative weight between diagonal and off-diagonal components is not fixed. We therefore introduce a universal real parameter $r$ and redefine
\begin{equation}
    A_f \equiv r\,D_f + O_f.
    \label{eq:Af-defined}
\end{equation}
The exceptional Jordan algebra $J_3(\mathbb{O}_\mathbb{C})$ is associated with the exceptional Lie groups $F_4$ and $E_6$, where $F_4$ acts as the automorphism group preserving the Jordan structure, while $E_6$ acts as the reduced structure group~\cite{FultonHarris, Baez:2001dm, Dray:1999xe}. In particular, the intrinsic invariants of the algebra are controlled by $F_4$. In the octonionic construction of Sec.~\ref{sec:fermion_generations}, fermion states arise as eigenstates of algebraic operators acting on $\mathbb{C}\otimes\mathbb{O}$, and an occupation-number grading is obtained from the associated number operator in the Clifford-algebra construction~\cite{Furey:2010fm, Furey:2015yxg}. This provides an algebraic organization of candidate state labels and an occupation-number grading, but does not by itself establish the complete Standard Model gauge quantum numbers or their compatibility with the full weak-isospin and hypercharge assignments. The embedding into $J_3(\mathbb{O}_\mathbb{C})$ introduces a complementary structure in which each fermion sector is described by a Hermitian element $A_f$ defined in Eq.~\eqref{eq:Af-defined}. The characteristic equation of $A_f$ defines three Jordan eigenvalues. In the real-invariant slices used in the phenomenological fit, these roots are real; in the fully complexified algebra, they can be complex, as reviewed in Appendix~\ref{app:J3OC}. The hierarchy construction below uses the ordered positive spectral scales formed from their magnitudes. We identify these spectral scales with the fundamental quantities governing fermion mass hierarchies, so that mass ratios arise directly from their relative structure. In this way, the occupation-number grading and the Jordan spectrum play distinct roles in the model: the former organizes candidate charge labels, while the latter provides the spectral quantities used in the phenomenological mass parametrization. Their identification with physical charges and masses relies on the additional assignments and mapping assumptions stated above. The octonionic Clifford structure organizes the candidate one-generation state space, while the exceptional Jordan embedding organizes inter-generation spectral structure. In this sense, the octonionic state-space construction and the Jordan spectral parametrization play distinct roles; their identification with physical gauge quantum numbers and flavour hierarchies depends on the additional assignments and mapping assumptions stated above. The parameter $r$ controls the relative normalization in the spectral parametrization, while the octonionic data $(x_f,y_f,z_f)$ determine the spectral splitting that generates the hierarchy. This identification provides the essential bridge between the algebraic construction and the observed fermion mass spectrum, and motivates the spectral mass-ratio relations derived in Sec.~\ref{sec:spectral_construction}.

\section{Constrained Jordan Embeddings}
\label{sec:constrained-embeddings}
The construction developed in Sec.~\ref{sec:hermitian_embedding} maps each fermion sector to a specific Hermitian element $A_f$, establishing the conceptual framework where mass scales mirror Jordan eigenvalues. However, before these algebraic invariants can be explicitly mapped to physical mass ratios, we must map out the exact geometry of the underlying parameter space. The purpose of the present section is to determine the residual moduli structure of these embeddings after the kinematic and algebraic constraints of our framework are strictly enforced. We first analyze the unconstrained orbit structure of $J_3(\mathbb{O}_{\mathbb{C}})$ under the action of $F_4^{\mathbb{C}}$, then study how the symmetric and coassociative conditions reduce the orbit space, and finally show how the rigid framework of Ref.~\cite{Singh:2025xxv} arises as a distinguished point inside the resulting moduli space.

\subsection{Generic Orbit Structure of \texorpdfstring{$J_3(\mathbb{O}_{\mathbb{C}})$}{J3(OC)}}
A general Hermitian element of the exceptional Jordan algebra takes the form
\begin{equation}
A = \begin{pmatrix}
        \alpha_1 & x_1 & x_2^\dagger \\
        x_1^\dagger & \alpha_2 & x_3 \\
        x_2 & x_3^\dagger & \alpha_3
  \end{pmatrix},
   \qquad
    \alpha_i\in\mathbb{C}, \quad x_i\in\mathbb{O}_{\mathbb{C}}.
\end{equation}
Since each complexified octonion contributes 16 real degrees of freedom, the total real dimension is 54. Equivalently $\dim_\mathbb{C} J_3(\mathbb{O}_\mathbb{C}) = 27$, which is the complex dimension of the fundamental representation $\mathbf{27}$ of $E_6^\mathbb{C}$. The natural symmetry group acting on $J_3(\mathbb{O}_\mathbb{C})$ is the complex automorphism group of the Jordan algebra,
\begin{equation}
    \mathrm{Aut}(J_3(\mathbb{O}_\mathbb{C})) \cong F_4^\mathbb{C}, \qquad \dim_\mathbb{C} F_4^\mathbb{C}=52,
\end{equation}
which preserves the Jordan product together with the spectral invariants $\operatorname{tr}(A)$, $\sigma(A)$, and $\det(A)$. The orbit of an element $A\in J_3(\mathbb{O}_{\mathbb{C}})$ under $F_4^{\mathbb{C}}$ is
\begin{equation}
\mathcal O_A = \{g\cdot A \mid g\in F_4^{\mathbb{C}}\},
\end{equation}
and consists of all Jordan elements related to $A$ by exceptional symmetry transformations. For a generic semisimple element, the stabilizer subgroup is $\mathrm{Spin}(8,\mathbb{C})$, whose complex dimension is $28$. The generic orbit dimension is therefore $52-28=24$. Consequently, the generic orbit space $J_3(\mathbb{O}_{\mathbb{C}})/F_4^{\mathbb{C}}$ is three complex dimensional and may be parameterized by the
Jordan eigenvalues,
\begin{equation}
    J_3(\mathbb{O}_\mathbb{C})/F_4^\mathbb{C} \;\cong\; \big\{(\lambda_1,\lambda_2,\lambda_3)\in\mathbb{C}^3\big\}/S_3,
    \label{eq:full_moduli}
\end{equation}
where the quotient by $S_3$ reflects permutation symmetry of
the spectrum.

\subsection{Constrained Symmetric Sector and the Rigid Benchmark Point}
Here, $\operatorname{Sc}$ denotes projection onto the scalar component, $\Re$ denotes the real part, and $\|x\|^2\equiv\operatorname{Sc}(xx^\dagger)$. The corresponding definitions are given in Appendix~\ref{app:J3OC}. The hierarchy construction does not use the full orbit space of $J_3(\mathbb{O}_{\mathbb C})$. Instead, the framework is restricted to a constrained symmetric sector. First, the diagonal entries are taken equal,
\begin{equation}
\alpha_1=\alpha_2=\alpha_3=s,
\end{equation}
so that
\begin{equation}
A=s\mathbf{1}_3+O,
\end{equation}
with $O$ traceless and purely off-diagonal. Second, the off-diagonal octonionic entries satisfy $\| x_1 \|^2 = \| x_2 \|^2 = \| x_3 \|^2$. Finally, the cubic octonionic contribution is required to vanish,
\begin{equation}
\tau=\Re(x_2(x_1x_3))=0.
\end{equation}
Under these conditions, the spectrum takes the symmetric form
\begin{equation}
\lambda_i=\{s-\delta,s,s+\delta\},
\end{equation}
with
\begin{equation}
\delta^2=\frac12 \operatorname{Sc}(O^2).
\end{equation}
Within this constrained sector, the rigid framework of Refs.~\cite{Singh:2025xxv, Bhatt:2021cpg} correspond to a specific calibration of the residual parameters:
\begin{enumerate}
    \item \emph{Chosen normalization} $\delta^2=3/8$. We impose $\|x_i\|^2=1/8$ as a normalization choice for the symmetric spectrum. This fixes the spread within the adopted benchmark; it is not derived here as a physical mass scale from exceptional symmetry alone.
    \item \emph{Square-root-mass quantization $s_f$.} The family-center parameters are fixed using the number-operator structure of the octonionic Clifford construction. In suitable units, the rigid framework assigns the square-root-mass values $s_d = 1$, $s_u = \frac23$, $s_e = \frac13$ and $s_\nu = 0$. The corresponding electric-charge assignments obtained from the Furey construction~\cite{Furey:2015yxg} are $q_d=\frac13$, $q_u=\frac23$, $q_e=1$, and $q_\nu=0$. The relation between the charged-lepton and down-quark assignments is imposed through the cross-sector prescription described in Sec.~\ref{sec:spectral_construction}; it is not derived from the charge assignment alone.
    \item \emph{Rigid scale} $r=1$. The diagonal and off-diagonal pieces are taken on equal footing, with no rescaling.
    \item \emph{Vanishing cubic invariant} $\tau=0$. On the chosen coassociative slice, this specifies the cubic-invariant constraint. Its connection to the charged-lepton assignment is part of the adopted phenomenological cross-sector prescription.
    \item \emph{Phenomenological minimality assumption}. Each generation is assigned to a specific monomial of $\mathrm{Sym}^3(\mathbf{3})$ via a discrete chain selection that minimizes the path length in the cubic weight diagram subject to monotonicity. This is a selection assumption rather than a consequence of the representation theory alone. We adopt this criterion as a principle of parsimony: it selects the shortest monotonic chain compatible with the assumed spectral ordering, without introducing additional intermediate assignments.
\end{enumerate}
Under these calibrations, ladder-selection, and cross-sector assumptions, Ref.~\cite{Singh:2025xxv} gives closed-form relations for the charged-sector ratios, which are collected in Appendix~\ref{app:tp}. These relations are conditional on the adopted inputs and assignments. Promoting $r$, $p$, and $\Phi_e$ to fitted quantities in the present work is therefore a genuine weakening of the parameter-free status of the rigid point, not an additional parameter-free prediction. The constrained sector described above specifies the class of admissible Jordan spectra entering the hierarchy construction. The remaining task is to determine how fermion mass ratios emerge from these ordered spectral scales and whether small deformations of the rigid point preserve the ladder structure while lowering the unpenalized global log-residual. In the next section, we show that the consistency of hierarchical composition organizes the spectrum into cubic ladder structures and forces the resulting hierarchy relations to take power-law form.

\section{Spectral Construction of Fermion Mass Hierarchies}
\label{sec:spectral_construction}

\subsection{Spectral Hierarchies and Power Laws}
The algebraic framework developed in Sec.~\ref{sec:hermitian_embedding} associates to each fermion sector $f\in\{\nu,e,u,d\}$ a Hermitian element
\begin{equation}
    A_f\in J_3(\mathbb{O}_{\mathbb{C}}).
\end{equation}
The purpose of this section is to show how the hierarchy of fermion masses can be constructed directly from the spectrum of this element. The spectrum of \(A_f\) is determined by the Jordan characteristic equation, as discussed in Appendix~\ref{app:J3OC},
\begin{equation}
    \lambda^3-\operatorname{tr(}A_f)\lambda^2+\sigma(A_f)\lambda-\det(A_f)=0 .
    \label{eq:eigen}
\end{equation}
In the real-invariant sectors used for the charged-fermion fit below, the three roots are real. More generally, the roots of the complexified Jordan characteristic equation may be complex, and the hierarchy construction then uses the ordered positive spectral scales obtained from their magnitudes,
\begin{equation}
    |\lambda_1^{(f)}|\le |\lambda_2^{(f)}|\le |\lambda_3^{(f)}|.
\end{equation}
Because the Jordan eigenvalues are algebraic invariants of the exceptional Jordan algebra, they depend only on the intrinsic structure of the element $A_f$ and therefore provide the natural spectral data from which fermion mass hierarchies can be constructed. Since the algebraic construction fixes no overall mass scale, only mass ratios are physically meaningful. Denoting the masses of the three generations by
\begin{equation}
    m_1<m_2<m_3 ,
\end{equation}
the fundamental observables are therefore the adjacent ratios
\begin{equation}
    \frac{m_2}{m_1},\qquad \frac{m_3}{m_2},
\end{equation}
which correspond to successive steps of the generation hierarchy.

\paragraph*{Contrasts and power laws.}
A hierarchy may be interpreted as a sequence of contrasts between positive quantities. For two scales $A,B>0$ we define
\begin{equation}
    R\equiv\frac{A}{B}, \qquad \Delta\equiv\log R .
\end{equation}
Successive ladder steps compose multiplicatively,
\begin{equation}
    R_{13}=R_{12}R_{23},
\end{equation}
or equivalently, logarithmic contrasts add,
\begin{equation}
    \Delta_{13}=\Delta_{12}+\Delta_{23}.
\end{equation}
Thus, the fermion hierarchy can be interpreted as a sequence of additive logarithmic contrasts built from underlying spectral scales. Suppose a physical mass scale is determined by a positive spectral quantity \(\Lambda_i\). Let \(\Lambda\) denote a positive function of the eigenvalues \(\lambda_i\), i.e. \(\Lambda_i = \Lambda_i(\lambda_1,\lambda_2,\lambda_3) > 0\). Since only ratios are physically relevant, their precise form may vary between sectors
\begin{equation}
    m=F(\Lambda).
\end{equation}
If mass ratios depend only on contrasts between spectral scales,
\begin{equation}
    \frac{F(\Lambda_1)}{F(\Lambda_2)}=\Phi\!\left(\frac{\Lambda_1}{\Lambda_2}\right).
\end{equation}
Consistency with ladder composition requires
\begin{equation}
    \Phi(xy)=\Phi(x)\Phi(y).
\end{equation}
Under mild regularity assumptions\footnote{Without minimal regularity conditions such as continuity or measurability, the multiplicative Cauchy equation admits pathological solutions that are not physically meaningful.}, the only solutions are power laws,
\begin{equation}
    \Phi(x)=x^p .
\end{equation}
Consequently, hierarchical relations between spectral scales and masses must take the form
\begin{equation}
    \frac{m(\Lambda_1)}{m(\Lambda_2)}=\left(\frac{\Lambda_1}{\Lambda_2}\right)^p.
\end{equation}
or redefining $p\to 2p$, we get,
\begin{equation}
    \sqrt\frac{m(\Lambda_1)}{m(\Lambda_2)}=\left(\frac{\Lambda_1}{\Lambda_2}\right)^p.
\end{equation}
For ordered spectral scales $\Lambda_1 < \Lambda_2$, consistency with the hierarchy $m_1 < m_2$ requires $p > 0$. Thus, consistency of multiplicative hierarchy composition naturally leads to power-law relations between spectral scales and fermion masses.

\subsection{Cubic Ladders in \texorpdfstring{$\mathrm{Sym}^3(\mathbf{3})$}{Sym3(3)}}
\paragraph*{Cubic channel.} Because the exceptional Jordan algebra possesses a fundamental cubic norm, the associated spectral data naturally organize into cubic algebraic structures. This motivates describing generation amplitudes in terms of the symmetric cubic representation $\mathrm{Sym}^3(\mathbf3)$. The Jordan spectrum is determined by a cubic characteristic equation. It is therefore natural to organize the generation amplitudes in the cubic representation
\begin{equation}
    \mathrm{Sym}^3(\mathbf{3}),
\end{equation}
whose structure is summarized in Appendix~\ref{app:Sym3}. This representation contains ten homogeneous cubic monomials
\begin{equation}
    a^3,\ a^2b,\ a^2c,\ ab^2,\ abc,\ ac^2,\ b^3,\ b^2c,\ bc^2,\ c^3 .
\end{equation}
constructed from three spectral scales and provides the minimal algebraic space in which cubic spectral data combine to produce generation amplitudes. We denote the ordered eigenvalues by
\begin{equation}
    a\equiv|\lambda_1|,\qquad b\equiv|\lambda_2|,\qquad c\equiv|\lambda_3|,
\end{equation}
so that $a\le b\le c$. The letters $a,b,c$ therefore represent ordered spectral scales derived from the Jordan eigenvalues rather than independent algebraic variables.

\begin{figure}[htbp]
\centering
\includegraphics[width=0.75\columnwidth]{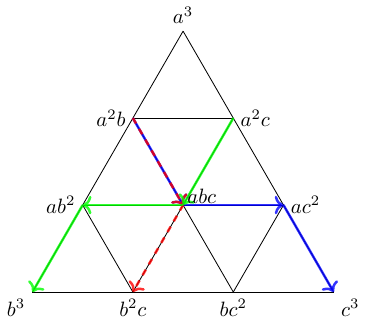}
\caption{Minimal ladders in the weight diagram of the symmetric cubic representation $\mathrm{Sym}^3(\mathbf{3})$, with ordered spectral letters $a \le b \le c$. Fermion mass hierarchies correspond to monotonic paths in which the multiplicity of the largest spectral scale $c$ increases relative to the smallest scale $a$. Colored arrows indicate the minimal ladders associated with different fermion sectors: down quarks (blue), up quarks (red), and charged leptons (green).}
\label{fig:cubic_ladder}
\end{figure}

\paragraph*{Cubic ladders.}
The representation $\mathrm{Sym}^3(\mathbf{3})$ and its ten cubic monomials are representation-theoretic objects. Their identification with physical fermion generations is an additional phenomenological assumption. In the present construction, we retain the minimality prescription of Ref.~\cite{Singh:2025xxv}: generation amplitudes are assigned to cubic monomials along the shortest admissible paths in the weight diagram, subject to the ordering $a\le b\le c$ and the monotonicity condition described below. This prescription is not implied uniquely by $\mathrm{Sym}^3(\mathbf{3})$ itself. Because $c$ represents the largest spectral scale while $a$ represents the smallest, hierarchical growth corresponds to monotonic paths in which the multiplicity of $c$ increases while that of $a$ decreases. In this geometric picture, the fermion hierarchy may therefore be viewed as an ordered monotonic flow on the weight diagram determined by the ordered spectral scales $a\le b\le c$. Different fermion sectors correspond to different minimal ladders terminating at different nodes of the cubic weight diagram, as shown in Fig.~\ref{fig:cubic_ladder}. Mass ratios are obtained by comparing neighboring amplitudes along these minimal ladders. Other admissible monotonic ladder assignments are possible in the same weight diagram and would, in general, lead to different phenomenological mass-ratio relations. Likewise, changes of basis can change the monomial representatives, although symmetry-related descriptions do not by themselves constitute distinct algebraic structures. We therefore do not claim uniqueness of the generation assignment from $\mathrm{Sym}^3(\mathbf{3})$ alone; the ladder assignment used here is retained as part of the phenomenological ansatz inherited from Ref.~\cite{Singh:2025xxv}.

\subsection{Fermion Hierarchy Relations}
\paragraph*{Quark sectors.}
For the quark sectors, the cubic invariant of the Jordan spectrum vanishes, implying that the magnitude of eigenvalues is symmetrically distributed about the central one,
\begin{equation}
    |\lambda_1|+|\lambda_3|=2|\lambda_2| .
\end{equation}
Although the eigenvalues are symmetric, the ladders generating the up- and down-quark hierarchies differ because different cubic amplitudes are associated with the physical generations. For down-type quarks, we adopt the ladder assignment
\begin{equation}
    \sqrt{m_d}\propto [a^2 b]^p,\qquad \sqrt{m_s}\propto [abc]^p,\qquad \sqrt{m_b}\propto [c^3]^p .
\end{equation}
The resulting relations are
\begin{align}
    \sqrt{\frac{m_b}{m_s}} = \left[\frac{c^2}{ab}\right]^p,
    \qquad
    \sqrt{\frac{m_s}{m_d}} = \left[\frac{c}{a}\right]^p.
\end{align}
For the up sector, we adopt the corresponding ladder terminating at the interior node $b^2c$,
\begin{equation}
    \sqrt{m_u}\propto [a^2 b]^p,\qquad \sqrt{m_c}\propto [abc]^p,\qquad \sqrt{m_t}\propto [b^2 c]^p .
\end{equation}
giving,
\begin{equation}
    \sqrt{\frac{m_t}{m_c}}=\left[\frac{b}{a}\right]^p.
\end{equation}
and
\begin{equation}
    \sqrt{\frac{m_c}{m_u}}=\left[\frac{c}{a}\right]^p.
\end{equation}

\paragraph*{Charged leptons.}
We relate the charged-lepton ladder to the down-quark ladder using a prescribed $b\leftrightarrow c$ reflection of the weight diagram of $\mathrm{Sym}^3(\mathbf{3})$. Here this reflection is a phenomenological mapping of the selected weight labels; we do not identify it with the outer automorphism of the $A_2$ Dynkin diagram.
\begin{figure}[htbp]
\centering
\includegraphics[width=0.75\columnwidth]{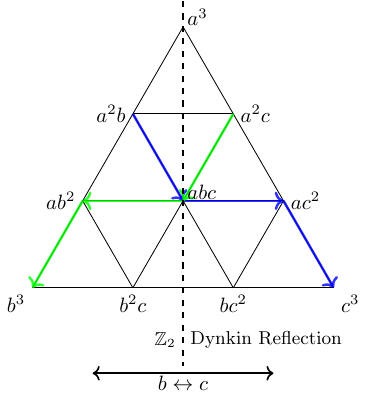}
\caption{Prescribed $b\leftrightarrow c$ reflection of the weight diagram of $\mathrm{Sym}^3(\mathbf{3})$. The reflection maps the selected down-quark ladder to the selected charged-lepton ladder while leaving the intermediate monomial $abc$ invariant. It is used here as a phenomenological weight-assignment prescription.}
\label{fig:dynkin-reflection}
\end{figure}
Applying this reflection to the down-sector ladder
\begin{equation}
a^2 b \;\rightarrow\; abc \;\rightarrow\; c^3
\end{equation}
produces the charged-lepton ladder
\begin{equation}
a^2 c \;\rightarrow\; abc \;\rightarrow\; b^3 .
\end{equation}
The prescribed reflection maps the selected down-quark monomials to the selected charged-lepton monomials, as illustrated in Fig.~\ref{fig:dynkin-reflection}. This mapping determines the chosen ladder assignment, but does not by itself imply a relation between the physical mass functionals in the two sectors. To obtain the cross-sector relation used in the present phenomenological analysis, we additionally impose a transport prescription that identifies the corresponding upper-rung ratio. Under this additional assumption, we use
\begin{equation}
    \sqrt{\frac{m_\tau}{m_\mu}}=\sqrt\frac{m_s}{m_d}.
\end{equation}
The lower-rung relation requires an additional phenomenological parameter describing the residual charged-lepton hierarchy, which we write as a spectral tilt,
\begin{equation}
    T_\ell=\left|\frac{\lambda^{(\ell)}_{\max}}{\lambda^{(\ell)}_{\min}}\right|,
\end{equation}
which measures the extremal asymmetry of the lepton spectrum and encodes the residual hierarchy beyond the cubic ladder. After factoring out the upper rung, the remaining contrast is controlled by this spectral tilt, reflecting the incomplete cancellation of the cubic structure under the Dynkin mapping,
\begin{equation}
    \sqrt{\frac{m_\mu}{m_e}} = \sqrt{\frac{m_\tau}{m_\mu}}\,T_\ell^p.
\end{equation}

\paragraph*{Neutrinos.}
Neutrinos exhibit a qualitatively different hierarchy. Experimentally, only mass-squared differences are measured, and neutrino masses are parametrically suppressed relative to charged fermions. Nevertheless, ladder consistency still applies. The minimal relation between spectral scales and neutrino masses, therefore, takes a power-law form. In this sector, we take the spectral quantities to be the magnitudes of the eigenvalues, \(\Lambda_i = |\lambda_i|\), so that the hierarchy is determined directly by the ordered spectral scales.
\begin{equation}
    \sqrt m_{i}\propto\Lambda_i^{p},
\end{equation}
where $\Lambda_i$ are positive spectral scales constructed from the Jordan eigenvalues and $p > 0$ is a universal exponent. The physical neutrino masses $(m_{\nu_1}, m_{\nu_2}, m_{\nu_3})$ are defined according to the standard oscillation convention and are not, in general, ordered by magnitude. Instead, they are related to the ordered spectral masses $(m_1, m_2, m_3)$ through an ordering-dependent identification. For normal ordering (NO), where $m_{\nu 1} < m_{\nu 2} < m_{\nu 3}$, we identify $(m_{\nu 1}, m_{\nu 2}, m_{\nu 3}) = (m_1, m_2, m_3)$. For inverted ordering (IO), where $m_{\nu 3} < m_{\nu 1} < m_{\nu 2}$, the identification becomes $(m_{\nu 1}, m_{\nu 2}, m_{\nu 3}) = (m_2, m_3, m_1)$. The resulting ratios for the ordered spectral masses are
\begin{align}
    \sqrt\frac{m_{3}}{m_{2}} = \left[\frac{\Lambda_3}{\Lambda_2}\right]^p,
    \qquad
    \sqrt\frac{m_{2}}{m_{1}} = \left[\frac{\Lambda_2}{\Lambda_1}\right]^p .
\end{align}

\paragraph*{Summary of observable ratios.}
The fermion hierarchy derived from the Jordan spectrum can therefore be summarized as:\\[2pt]
\noindent \textbf{Down quarks:}
\begin{align}
    \sqrt{\frac{m_b}{m_s}}=\left[\frac{c^2}{ab}\right]^p,\qquad \sqrt{\frac{m_s}{m_d}}=\left[\frac{c}{a}\right]^p
    \label{eq:down-ratio}
\end{align}
\noindent \textbf{Up quarks:}
\begin{align}
    \sqrt{\frac{m_t}{m_c}}=\left[\frac{b}{a}\right]^p,\qquad \sqrt{\frac{m_c}{m_u}}=\left[\frac{c}{a}\right]^p
    \label{eq:up-ratio}
\end{align}
\noindent \textbf{Charged leptons:}
\begin{align}
    \sqrt{\frac{m_\tau}{m_\mu}}=\sqrt{\frac{m_s}{m_d}},\qquad \sqrt{\frac{m_\mu}{m_e}}=\sqrt{\frac{m_\tau}{m_\mu}}\,T_\ell^p
    \label{eq:electron-ratio}
\end{align}
\noindent \textbf{Neutrinos:}
\begin{align}
    \sqrt\frac{m_{3}}{m_{2}}=\left[\frac{\Lambda_3}{\Lambda_2}\right]^p,\qquad \sqrt\frac{m_{2}}{m_{1}}=\left[\frac{\Lambda_2}{\Lambda_1}\right]^p
    \label{neutrino-ratio}
\end{align}

\section{Phenomenology of Mass Hierarchies}
\label{sec:results}

\subsection{Charged-Fermion Sector}
The fermion mass hierarchy in the present framework is determined by the Jordan eigenvalues of the Hermitian element $A_f\in J_3(\mathbb{O}_{\mathbb{C}})$ introduced in Sec.~\ref{sec:spectral_construction}. These eigenvalues are obtained from Eq.~\eqref{eq:eigen}. The spectral invariants entering this equation depend only on three intrinsic quantities that characterize the octonionic family structure. The first quantity is the family center $s$, which appears in the diagonal part of the matrix $A_f$. Physically, this parameter sets the common spectral origin around which the three eigenvalues are distributed. In the present framework, $s$ is closely related to the algebraic charge assignment emerging from the Clifford-octonionic construction. In particular, the ordering of fermion sectors follows a reversed charge hierarchy relative to the electric charge ratios $(1:2:3)$ appearing in the occupation number spectrum. Under the sector-assignment prescription adopted here, this ordering becomes $(3:2:1)$, a feature emphasized in Refs.~\cite{Singh:2025xxv, Bhatt:2021cpg}. The second invariant,
\begin{equation}
    \delta^2 = \operatorname{Sc}(x x^\dagger) + \operatorname{Sc}(y y^\dagger) + \operatorname{Sc}(z z^\dagger) ,
\end{equation}
measures the magnitude of the inter-generation octonionic components appearing in Eq.~\eqref{eq:states}. Here $\operatorname{Sc}(\cdot)$ denotes projection onto the scalar part of $\mathbb{C}\otimes\mathbb{O}$, as defined in Appendix~\ref{app:J3OC}. This quantity controls the overall separation of the three eigenvalues and therefore determines the basic scale of generation splitting. The third invariant,
\begin{equation}
    \tau=\Re(y(xz)),
\end{equation}
encodes the octonionic triple product. Unlike $\delta$, which controls the magnitude of the splitting, $\tau$ introduces an asymmetry among the eigenvalues and therefore determines the tilt of the spectrum. These relations follow directly from the Jordan eigenvalue problem and the definition of scalar projection in Appendix~\ref{app:J3OC}, with the resulting invariant parameters for each sector summarized in Table~\ref{tab:invariants}.
\begin{table}[htbp]
\centering
\setlength{\tabcolsep}{3pt}
\renewcommand{\arraystretch}{1.2}
\begin{tabular}{cccc}
\hline\hline
Sector & $s$ & $\delta$ & $\tau$ \\
\hline
$d$   & $1$ & $\sqrt{\frac{3}{8}}$ & $0$ \\
$u$   & $\frac{2}{3}$ & $\sqrt{\frac{3}{8}}$ & $0$ \\
$e$   & $\frac{1}{3}$ & $\sqrt{\frac{3}{8}}$ & $\frac{1}{64}\cos\Phi_e$ \\
$\nu$ & $0$ & $\sqrt{\frac{3}{8}}$ & $\frac{1}{64}\cos\Phi_\nu$ \\
\hline\hline
\end{tabular}
\caption{Invariant parameters $(s,\tau,\delta)$ for the four fermion sectors entering the Jordan characteristic equation. The phases are defined as $\Phi_\nu\equiv\sum_a \alpha_{a,\nu}$ and $\Phi_e\equiv\sum_a \alpha_{a,e}$.}
\label{tab:invariants}
\end{table}

Using the invariant parameters from Table~\ref{tab:invariants}, we define the general Hermitian element as
\begin{equation}
    A(rs;x,y,z)=
                \begin{pmatrix}
                    rs & x & y^\dagger\\
                    x^\dagger & rs & z\\
                    y & z^\dagger & rs
                \end{pmatrix},
\end{equation}
the corresponding spectral invariants are
\begin{align}
    \operatorname{tr}(A)&=3rs,\\
    \sigma(A)&=3r^2s^2-\delta^2,\\
    \det(A)&=r^3s^3-rs\delta^2+2\tau .
\end{align}
The three eigenvalues can then be written in closed form using the trigonometric solution of the cubic equation
\begin{equation}
    \lambda_k=rs+\frac{2\delta}{\sqrt{3}}\cos\!\left(\theta-\frac{2\pi k}{3}\right),\qquad k=0,1,2 ,
    \label{eq:eigenval}
\end{equation}
where
\begin{equation}
    \cos(3\theta)=\frac{3\sqrt{3}\tau}{\delta^3}.
\end{equation}
Ordering the solutions as
\begin{equation}
    a\equiv|\lambda_1|\le b\equiv|\lambda_2|\le c\equiv|\lambda_3| ,
\end{equation}
defines the three spectral scales entering the cubic ladder relations from Eqs.~\eqref{eq:down-ratio}--\eqref{neutrino-ratio}. All observable fermion mass ratios are therefore determined by the invariant parameters $(s,\delta,\tau)$ through the ordered magnitude of eigenvalues $(a,b,c)$.

We now determine the charged fermion mass ratios within the spectral framework. All observable ratios are functions of the ordered eigenvalues $(a,b,c)$, which are in turn fixed by the invariant parameters $(s,\delta,\tau)$ through Eq.~\eqref{eq:eigenval}. In the charged fermion sector, the adjustable parameters are $(r,p,\Phi_e)$, which control the relative normalization, the hierarchy exponent, and the charged-lepton spectral asymmetry of the underlying Jordan element. These three quantities are determined by minimizing a global logarithmic squared error over the six adjacent charged-fermion ratios,
\begin{equation}
\chi^2_{\log} = \sum_i \left[ \log R_i^{\text{th}} - \log R_i^{\text{exp}} \right]^2,
\end{equation}
which is consistent with the multiplicative structure of the hierarchy derived in Sec.~\ref{sec:spectral_construction}. The minimization yields the best-fit values $r=-0.98747$, $p=0.98747$, $\cos\Phi_e=-0.50877$, with unpenalized logarithmic squared error $\chi^2_{\log} = 0.0745$. The corresponding rigid values from Ref.~\cite{Singh:2025xxv}, evaluated on the same six central experimental ratios, give $\chi^2_{\log}\simeq0.106$. Thus, the fitted deformation lowers the unpenalized global log-residual. This statement should not be confused with a parameter-count-adjusted preference: the present fit uses three continuous quantities for six ratios, whereas the rigid charged-sector formulae of Ref.~\cite{Singh:2025xxv} use none. For example, using the simple unweighted Gaussian information-criterion proxy \(\mathrm{AIC}=n\log(\chi^2_{\log}/n)+2k\) and \(\mathrm{BIC}=n\log(\chi^2_{\log}/n)+k\log n\) with \(n=6\), the rigid point gives \(\mathrm{AIC}_{\rm rigid}\simeq -24.22\) and \(\mathrm{BIC}_{\rm rigid}\simeq -24.22,\) while the deformed fit gives \(\mathrm{AIC}_{\rm def}\simeq -20.33\) and \(\mathrm{BIC}_{\rm def}\simeq -20.96.\) Thus \(\Delta \mathrm{AIC}\simeq +3.88\) and \(\Delta \mathrm{BIC}\simeq +3.26,\) so the parameter penalty does not favor the deformed fit.

A notable outcome is that the best-fit hierarchy exponent satisfies $p \simeq 1$, indicating proximity to square-root spectral scaling. The fit also yields numerically similar values for $|r|$ and $p$. This near equality is not imposed and is not explained dynamically here. It may indicate a hidden constraint or a parameter degeneracy; a covariance analysis or a constrained fit with $r=-p$ would be required before treating $r$ and $p$ as independently established moduli. All experimental mass ratios are constructed from running masses evaluated at a common scale $\mu = M_Z$, ensuring consistency across fermion sectors. The choice of $\mu=M_Z$ is motivated by the parent octonionic-unification programme, which hypothesizes that the transition from a pre-spacetime, pre-quantum regime to the observed spacetime and quantum regime occurs around electroweak symmetry breaking. In that picture, the unified interaction separates into the electroweak, QCD, and gravitational sectors around this scale. This is a motivating hypothesis of the parent programme, not a dynamical result derived in the present spectral analysis. We therefore use $M_Z$ as the reference scale for the phenomenological input. The present framework does not specify renormalization-group evolution of the spectral moduli or Yukawa couplings, and hence makes no prediction for the corresponding mass ratios at the Planck scale. Such a comparison would require an RG-improved extension of the framework. The results are summarized in Table~\ref{tab:observed}.
\begin{table*}[htbp]
\centering
\setlength{\tabcolsep}{3pt}
\begin{tabular}
{
c
S[table-format=2.5(1.5)]
S[table-format=2.4]
S[table-format=2.2]
S[table-format=2.5]
S[table-format=2.2]
}
\hline\hline
{Observable} & {Experiment} & {Ref.~\cite{Singh:2025xxv}} & {Err. (\%)} & {Present Work} & {Err. (\%)} \\
\hline
$\sqrt{m_b/m_s}$ & 7.24882 +- 1.02764 & 6.70681 & 7.48 & 6.74478 & 6.95 \\
$\sqrt{m_s/m_d}$ & 4.35494 +- 1.09947 & 4.15959 & 4.49 & 4.18832 & 3.83 \\
$\sqrt{m_t/m_c}$ & 16.65482 +- 1.13938 & 12.27878 & 26.27 & 13.85963 & 16.78 \\
$\sqrt{m_c/m_u}$ & 22.07717 +- 4.26963 & 23.55755 & 6.71 & 26.53252 & 20.18 \\
$\sqrt{m_\tau/m_\mu}$ & 4.12314 +- 0.00023 & 4.15959 & 0.88 & 4.18832 & 1.58 \\
$\sqrt{m_\mu/m_e}$ & 14.52951 +- 0.00000 & 14.09749 & 2.97 & 15.19412 & 4.57 \\
\hline\hline
\end{tabular}
\caption{Comparison of fermion mass ratios between experiment, Ref.~\cite{Singh:2025xxv}, and the present work. The column ``Ref.~\cite{Singh:2025xxv}'' lists the predictions obtained in the framework of Singh, while ``Present Work'' denotes the results derived in the present spectral construction. Experimental inputs are taken from Tables II and III of Ref.~\cite{Xing:2007fb}, using running masses evaluated at the common scale $\mu = M_Z$, from which the ratios are constructed; the quoted uncertainties correspond to $1\sigma$ values. The quoted error (Err.~(\%)) represents the percentage deviation from the experimental central values.}
\label{tab:observed}
\end{table*}
Table~\ref{tab:observed} shows a mixed pattern. The fitted deformation slightly improves the two down-sector ratios and substantially reduces the large discrepancy in $\sqrt{m_t/m_c}$. However, $\sqrt{m_c/m_u}$ and both charged-lepton ratios are farther from their central experimental values than in the rigid benchmark. The correct quantitative statement is therefore that the unpenalized global logarithmic error is reduced, not that every charged-sector ratio is improved or that the up-quark and charged-lepton sectors are reproduced at the sub-percent level.

The distinction from Ref.~\cite{Singh:2025xxv} is nevertheless structurally useful. The present construction relaxes the relative normalization between the diagonal and off-diagonal components of $A_f$ through $r$, allows a universal hierarchy exponent $p$ rather than fixing the square-root scaling exactly, and retains the charged-lepton octonionic triple-product phase through $\tau = \Re(y(xz))$. Among these changes, the nonzero octonionic phase is the genuinely new spectral ingredient; the parameters $r$ and $p$ are phenomenological relaxations of the rigid point. Future fits should isolate these effects one at a time to determine which deformation actually drives the reduction in global log-residual. For completeness, the corresponding closed-form expressions of Ref.~\cite{Singh:2025xxv} are listed in Appendix~\ref{app:tp}.

\subsection{Neutrino Sector}
Unlike the charged-fermion sector, current neutrino data determine only mass-squared differences and mixing parameters rather than the full absolute mass spectrum. Consequently, the spectral framework constrains primarily the hierarchy structure rather than unique neutrino masses. The present neutrino discussion is deliberately more modest than that of Ref.~\cite{Singh:2025xxv}: we do not derive a Majorana condition, a PMNS mixing pattern, or a leptonic CP phase prediction here. The goal is only to show that the same spectral language is compatible with both standard mass orderings and with the cosmological bound on the total neutrino mass. In this case, the universal exponent $p$ is fixed entirely from the charged-fermion sector and is therefore not treated as an additional free parameter. Unlike charged fermions, only neutrino mass-squared differences are experimentally measured, and the absolute neutrino mass scale remains unknown. We therefore focus on the hierarchy structure implied by the spectral construction itself.

The experimentally measured mass-squared differences, taken from the NuFIT 6.1 global analysis~\cite{NuFIT6.1}, are
\begin{align}
\Delta m^2_{\rm sol} &= m_{\nu_2}^2 - m_{\nu_1}^2\simeq 7.537^{+0.094}_{-0.10}\times10^{-5}\ {\rm eV}^2.
\end{align}
\begin{table}[htbp]
\centering
\setlength{\tabcolsep}{3pt}
\renewcommand{\arraystretch}{1.2}
\begin{tabular}{lcc}
\hline\hline
Ordering & Definition & Value \\
\hline
Normal (NO) & $m_{\nu_3}^2 - m_{\nu_1}^2$ & $+2.511^{+0.021}_{-0.020} \times 10^{-3}\ \text{eV}^2$ \\
Inverted (IO) & $m_{\nu_3}^2 - m_{\nu_2}^2$ & $-2.483^{+0.020}_{-0.020} \times 10^{-3}\ \text{eV}^2$ \\
\hline\hline
\end{tabular}
\caption{Atmospheric mass-squared differences for normal and inverted neutrino mass ordering. Experimental values are taken from the NuFIT 6.1 global analysis~\cite{NuFIT6.1}, using the IC24 dataset without SK atmospheric data, and correspond to best-fit values with $1\sigma$ uncertainties.}
\label{tab:atm_splitting}
\end{table}
The atmospheric mass-squared differences for the two orderings are summarized in Table~\ref{tab:atm_splitting}. Unlike the charged fermion case, no common renormalization scale is required here, since the analysis is based on mass-squared differences rather than individual running masses. In the present framework described in Sec.~\ref{sec:spectral_construction}, the spectral construction yields ordered eigenvalues
\begin{align}
    a \le b \le c,
\end{align}
which defines intrinsic hierarchy scales from which neutrino masses arise through the universal power-law relation
\begin{align}
\sqrt m_i \propto \Lambda_i^{\,p},
\end{align}
where $\Lambda_i$ are positive spectral quantities constructed from the Jordan eigenvalues. Motivated by the symmetric structure of the cubic Jordan spectrum, we consider the neutrino sector to satisfy an approximately reflection-symmetric spectral configuration about the central eigenvalue. Consequently, the constrained spectral structure is consistent with configurations containing one parametrically suppressed state together with a comparatively compressed pair.

For normal ordering,
\begin{align}
m_{\nu_1} < m_{\nu_2} < m_{\nu_3},
\end{align}
the spectral hierarchy implies
\begin{align}
m_{\nu_1} \ll m_{\nu_2} < m_{\nu_3},
\end{align}
so that the lightest neutrino state becomes parametrically suppressed relative to the heavier two states. Using the oscillation data,
\begin{align}
m_{\nu_2} \simeq \sqrt{\Delta m^2_{\rm sol}}, \qquad
m_{\nu_3} \simeq \sqrt{\Delta m^2_{\rm atm}},
\end{align}
while $m_{\nu_1}$ remains parametrically small.

For inverted ordering,
\begin{align}
m_{\nu_3} < m_{\nu_1} < m_{\nu_2},
\end{align}
the spectral construction is consistent with a quasi-degenerate pair,
\begin{align}
m_{\nu_1} \simeq m_{\nu_2}
\simeq \sqrt{|\Delta m^2_{\rm atm}|},
\end{align}
together with a parametrically suppressed lightest state $m_{\nu_3}$. Both orderings may therefore be interpreted within the same spectral framework, differing primarily in the identification between ordered spectral scales and physical neutrino states.

For a parametrically suppressed lightest state, the corresponding total neutrino masses are approximately $\sum_i m_{\nu_i} \approx 0.059\,\mathrm{eV}$ for normal ordering and $\sum_i m_{\nu_i} \approx 0.10\,\mathrm{eV}$ for inverted ordering, both remaining below the current cosmological bound $\sum_i m_{\nu_i}\lesssim 0.12\,\mathrm{eV}$~\cite{Planck:2018vyg}. The neutrino sector, therefore, remains less constrained than the charged-fermion sector within the present framework. Nevertheless, the spectral construction accommodates both normal and inverted ordering while maintaining consistency with oscillation data and current cosmological limits on the total neutrino mass.

\section{Conclusion and Outlook}
\label{sec:conclusion}

\subsection{Deformation of the Rigid Framework}
The rigid framework of Refs.~\cite{Singh:2025xxv, Bhatt:2021cpg} corresponds to a distinguished point in the constrained Jordan sector, obtained by fixing the normalization, hierarchy exponent, and octonionic phase structure entering the spectral construction. In the present work, these quantities are instead treated as effective phenomenological parameters determined through a global fit to the charged-fermion hierarchy data. The resulting best-fit values,
\begin{equation}
r \simeq -0.987, \qquad
p \simeq 0.987, \qquad
\cos\Phi_e \simeq -0.509,
\end{equation}
lie close to the square-root hierarchy limit. In particular, the near-unity value of the universal hierarchy exponent,
\begin{equation}
p \simeq 1,
\end{equation}
indicates that the observed charged-fermion hierarchies are close to square-root spectral scaling. The near numerical equality $|r|\simeq p$ remains unexplained and should be treated as a fit diagnostic rather than as a derived relation. From this viewpoint, the present framework is a phenomenological deformation of the rigid exceptional-Jordan construction, in which the relative normalization, exponent, and octonionic phase data are promoted to effective spectral moduli constrained by data rather than fixed \emph{a priori}. The underlying cubic ladder structure, Dynkin reflection symmetry, and minimal chain assignments remain unchanged.

\subsection{Summary and Phenomenology}
In this work, we have developed a spectral framework for fermion mass hierarchies based on the exceptional Jordan algebra $J_3(\mathbb{O}_\mathbb{C})$. Starting from an octonionic state-space construction in $\mathbb{C}\otimes\mathbb{O}$, the selected family-labelled states were embedded into Hermitian elements $A_f\in J_3(\mathbb{O}_\mathbb{C})$, whose eigenvalues define three ordered spectral scales $(a,b,c)$. These spectral invariants, protected by the automorphism group $F_4$, provide the fundamental quantities from which fermion mass ratios are constructed. A central result of this work is that fermion mass hierarchies can be organized in terms of cubic ladder structures in the symmetric representation $\mathrm{Sym}^3(\mathbf{3})$. Consistency of hierarchical composition enforces power-law relations between masses and spectral scales.

In the charged fermion sector, the mass ratios are described in terms of the fitted parameter set $(r,p,\Phi_e)$, which controls the relative normalization, hierarchy exponent, and charged-lepton octonionic phase structure of the Jordan element. Fixing these parameters via a global logarithmic fit yields a lower unpenalized global log-residual than the rigid point of Ref.~\cite{Singh:2025xxv}. This reduction is not uniform across individual observables: the down-sector ratios and $\sqrt{m_t/m_c}$ improve, while $\sqrt{m_c/m_u}$ and the charged-lepton ratios worsen relative to the rigid benchmark. The present work should therefore be read as a fitted spectral deformation that preserves the algebraic ladder architecture, not as a parameter-free replacement for Ref.~\cite{Singh:2025xxv}.

In the neutrino sector, the framework allows two hierarchy patterns corresponding to normal and inverted mass ordering. This part of the analysis is a compatibility statement rather than a sharp prediction of the absolute spectrum or of mixing. In particular, the lightest neutrino mass can be parametrically suppressed, while the total neutrino mass remains within current cosmological bounds. A derivation of Majorana structure, PMNS mixing, or leptonic CP violation is outside the scope of the present work.

\subsection{Dynamical Outlook and Exceptional Geometry}
The exceptional Jordan algebra plays a structural role in the present framework, organizing fermion hierarchy patterns through spectral invariants and cubic ladder relations rather than arbitrary Yukawa couplings. At the same time, the mapping from octonionic fermion states to Hermitian Jordan elements, together with the origin of the effective parameters $(r,\Phi_e,\Phi_\nu,p)$, remains an open dynamical problem.

The present framework should be interpreted as an effective low-energy spectral description of fermion hierarchies rather than as a fundamental ultraviolet theory. All charged-fermion ratios used in Sec.~\ref{sec:results} are evaluated at the common reference scale $\mu=M_Z$, which is close to the electroweak-symmetry-breaking regime and is the scale at which the running-mass inputs used here are specified. The exceptional Jordan structure introduced here is therefore intended to organize emergent low-energy spectral data after symmetry breaking and coarse-graining. CKM mixing, PMNS mixing, and renormalization-group running of the fitted spectral moduli are not addressed in this paper.

Possible underlying frameworks such as Trace Dynamics~\cite{Farnsworth:2026aae, Lochan:2011nr} may operate at a more fundamental pre-quantum level, while ordinary quantum field theory and renormalization-group flow emerge only in the effective low-energy description. The present work does not attempt to derive the running of Yukawa couplings, but instead focuses on the algebraic structure of the resulting electroweak-scale mass hierarchies.

Overall, the results suggest that fermion mass hierarchies may reflect an underlying exceptional algebraic structure, with the spectral properties of $J_3(\mathbb{O}_{\mathbb C})$ providing a common organizing language across fermion sectors. Further dynamical and geometric interpretations of the present framework, including a Spin(9)-adapted analysis of the cubic Jordan invariant, are discussed in Appendices~\ref{app:geometric} and~\ref{app:spin9_peirce}.

\begin{acknowledgments}
B.\,G.\,T. would like to thank Dawood Kothawala, Department of Physics, IIT Madras, for his guidance and support as co-supervisor, and for enabling this work to be carried out as part of the M.Sc. project. He also thanks the Inter-University Centre for Astronomy and Astrophysics (IUCAA), Pune, for hospitality and accommodation during a research visit.
\end{acknowledgments}

\section*{Data Availability Statement}
This manuscript has no associated data. All input values used in this work (the running fermion masses at $\mu = M_Z$, the neutrino oscillation parameters, and the cosmological bound on $\sum_i m_{\nu_i}$) are publicly available in Refs.~\cite{Xing:2007fb, NuFIT6.1, Planck:2018vyg}, and all derived results are reported within the article.

\section*{Code Availability Statement}
The numerical fit of the spectral moduli $(r, p, \Phi_e)$ and the neutrino-sector calculations were performed with short custom scripts, which are available from the corresponding author on reasonable request.

\appendix

\section{Octonions}
\label{app:octonions}
\begin{figure}[htbp]
\centering
\includegraphics[width=0.75\columnwidth]{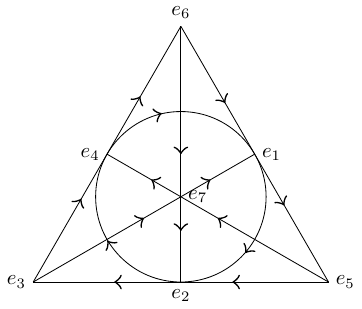}
\caption{Fano plane encoding octonion multiplication. Each oriented line represents a quaternionic triple. Cyclic order determines the sign convention.}
\label{fig:fano}
\end{figure}

The octonions $\mathbb{O}$ are the largest normed division algebra over $\mathbb{R}$. Together with $\mathbb{R}$, $\mathbb{C}$, and $\mathbb{H}$, they complete the sequence of real normed division algebras~\cite{Baez:2001dm}. An octonion $x\in\mathbb{O}$ can be written as
\begin{equation}
    x = x_0e_0 + \sum_{i=1}^{7} x_i e_i,    \qquad   x_\mu \in \mathbb{R},
\end{equation}
where $e_0 \equiv 1$ is the identity and $\{e_1,\dots,e_7\}$ are imaginary units satisfying
\begin{equation}
    e_i^2 = -1, \qquad i=1,\dots,7.
\end{equation}
Multiplication among the imaginary units is encoded by the oriented Fano plane shown in Fig.~\ref{fig:fano}.
Each oriented line in the diagram corresponds to a quaternionic triple. If $(e_i,e_j,e_k)$ lie on an oriented line, then
\begin{equation}
    e_i e_j = e_k, \qquad e_j e_k = e_i, \qquad e_k e_i = e_j,
\end{equation}
and reversing the order introduces a minus sign. Every pair of imaginary units, therefore, generates a quaternionic subalgebra.

Unlike the quaternions, the octonions are not associative. In general,
\begin{equation}
    (xy)z \neq x(yz).
\end{equation}
The quadratic norm is defined by
\begin{equation}
    |x|^2 = x \overline{x} = x_0^2 + \sum_{i=1}^{7} x_i^2,
\end{equation}
where conjugation is given by
\begin{equation}
    \overline{x} = x_0 - \sum_{i=1}^{7} x_i e_i.
\end{equation}
The norm satisfies $|xy|=|x||y|$, ensuring that $\mathbb{O}$ is a normed division algebra. This definition applies to real octonions. In the complexified algebra $\mathbb{C}\otimes\mathbb{O}$, the norm is no longer positive definite. The automorphism group of the octonions,
\begin{equation}
    G_2 = \mathrm{Aut}(\mathbb{O}),
\end{equation}
is a 14-dimensional exceptional Lie group. Fixing a chosen imaginary unit $e_a$ selects an $SU(3)$ subgroup preserving that direction. In this work, we use the complexified octonions
\begin{equation}
    \mathbb{O}_{\mathbb{C}}=\mathbb{C}\otimes\mathbb{O},
\end{equation}
whose elements are
\begin{equation}
    x = \sum_{\mu=0}^{7} z_\mu e_\mu, \qquad z_\mu \in \mathbb{C}.
\end{equation}
Complex scalars commute with the octonionic units, while the multiplication rules among the $e_i$ remain unchanged. This complexification permits Hermitian matrix constructions while preserving the underlying octonionic structure.

For complex octonions, it is important to distinguish three types of conjugation. First, complex conjugation, denoted by $^*$, acts only on the complex scalar, sending $i \mapsto -i$ while leaving the octonionic basis elements unchanged. Second, octonionic conjugation, denoted by an overbar, acts as $\overline{e_0} = e_0$ and $\overline{e_i} = -e_i$ for $i=1,\dots,7$. Finally, the Hermitian conjugation, denoted by $^\dagger$, combines both operations, so that for $x \in \mathbb{C}\otimes\mathbb{O}$ one has $x^\dagger = \overline{x^*} = (\overline{x})^*$.

\section{Spectral Structure of \texorpdfstring{$J_3(\mathbb{O}_{\mathbb{C}})$}{J3OC}}
\label{app:J3OC}
We now describe the spectral invariants of $J_3(\mathbb{C}\otimes\mathbb{O})$ that determine the eigenvalue structure used in the main text. The exceptional Jordan algebra, also known as the Albert algebra, is the algebra of $3\times3$ Hermitian matrices over the octonions~\cite{Baez:2001dm, Dray:1999xe, Springer2000, Jacobson}. In the real case, it is defined by
\begin{equation}
    J_3(\mathbb{O})=\{\mathcal{A}\in M_3(\mathbb{O}) \mid \mathcal{A}^\dagger=\bar {\mathcal{A}}^{\,T}=\mathcal{A}\},
\end{equation}
where $\bar{\phantom{x}}$ denotes octonionic conjugation. A generic element has the form
\begin{equation}
    \mathcal{A} = \begin{pmatrix}
            \alpha_1 & x_1 & \bar{x}_2 \\
            \bar{x}_1 & \alpha_2 & x_3 \\
        x_2 & \bar{x}_3 & \alpha_3
        \end{pmatrix},
        \qquad
        \alpha_i \in \mathbb{R},\; x_i \in \mathbb{O}.
\end{equation}
Since matrix multiplication over $\mathbb{O}$ is not associative, $J_3(\mathbb{O})$ is not closed under ordinary matrix products. Instead, one introduces the Jordan product
\begin{equation}
    \mathcal{A}\circ \mathcal{B}=\tfrac12(\mathcal{A}\mathcal{B}+\mathcal{B}\mathcal{A}),
\end{equation}
which is commutative and satisfies the Jordan identity
\begin{equation}
(\mathcal{A}\circ \mathcal{B})\circ \mathcal{A}^2=\mathcal{A}\circ(\mathcal{B}\circ \mathcal{A}^2).
\end{equation}
With this product, $J_3(\mathbb{O})$ becomes the unique simple formally real Jordan algebra possessing a cubic norm~\cite{Springer2000, Jacobson}. This algebra plays an important role in the theory of exceptional Lie groups.

In the real Albert algebra, the diagonal entries are real scalars, and the trace is simply the matrix trace,
\begin{equation}
    \operatorname{tr}(\mathcal{A})=\alpha_1+\alpha_2+\alpha_3.
\end{equation}
However, when forming Jordan powers such as $\mathcal{A}^2=\mathcal{A}\circ \mathcal{A}$ and $\mathcal{A}^3=\mathcal{A}\circ(\mathcal{A}\circ \mathcal{A})$, non-associative octonionic products appear. Although intermediate diagonal entries may contain octonionic components, all polynomial invariants of $J_3(\mathbb{O})$ are scalar-valued~\cite{Dray:1999xe, Springer2000}. In particular, the cubic norm involves the scalar part of triple octonionic products.

In the present work we employ the complexified algebra $J_3(\mathbb{O}_{\mathbb{C}})$, whose entries lie in $\mathbb{O}_{\mathbb{C}}=\mathbb{C}\otimes\mathbb{O}$. Complexification preserves all polynomial identities since these are algebraic relations over the base field. Let
\begin{equation}
    \operatorname{Sc}:\mathbb{O}_{\mathbb{C}}\to\mathbb{C}
\end{equation}
denote projection onto the scalar component (coefficient of $e_0$). Because Jordan products may generate non-scalar diagonal entries, the trace must be defined via scalar projection,
\begin{equation}
    \operatorname{tr}(\mathcal{A}) = \sum_{i=1}^3 \operatorname{Sc}(\mathcal{A}_{ii}),
\label{eq:scalartrace_app2}
\end{equation}
so that $\operatorname{tr}(\mathcal{A})\in\mathbb{C}$. This extension is natural: in the real algebra, the diagonal entries are already scalar, while in the complexified case, projection ensures that the trace remains valued in the base field and that the quadratic and cubic invariants remain well-defined~\cite{Springer2000}. The quadratic invariant is defined by
\begin{equation}
    \sigma(\mathcal{A}) = \frac{1}{2}\Big( (\operatorname{tr} \mathcal{A})^2 - \operatorname{tr}(\operatorname{\mathcal{A}}^2) \Big),
\end{equation}
where $\mathcal{A}^2 = \mathcal{A} \circ \mathcal{A}$ is the Jordan square. For a $\dagger$-Hermitian matrix, the scalar projection ensures that $\operatorname{tr}(\mathcal{A}^2)$ is well-defined and valued in $\mathbb{C}$. Explicitly, for
\begin{equation}
    \mathcal{A} =
                \begin{pmatrix}
                    \alpha_1 & x_1 & x_2^\dagger \\
                    x_1^\dagger & \alpha_2 & x_3 \\
                    x_2 & x_3^\dagger & \alpha_3
                \end{pmatrix},
\end{equation}
one obtains
\begin{align}
    \sigma(\mathcal{A}) &= \alpha_1 \alpha_2 + \alpha_2 \alpha_3 + \alpha_3 \alpha_1\\
              &- \Big( \operatorname{Sc}(x_1 x_1^\dagger) + \operatorname{Sc}(x_2 x_2^\dagger) + \operatorname{Sc}(x_3 x_3^\dagger) \Big).
\end{align}
In the main text, the off-diagonal entries $(x_f, y_f, z_f)$ correspond to $(x_1, x_2, x_3)$. The cubic invariant (determinant) is defined by
\begin{equation}
    \det(\mathcal{A}) = \frac{1}{3}\operatorname{tr}\big( (\mathcal{A}^\#) \circ \mathcal{A} \big),
\end{equation}
where $\mathcal{A}^\# = \mathcal{A} \circ \mathcal{A} - (\operatorname{tr}\mathcal{A})\mathcal{A} + \sigma(\mathcal{A})\mathbf{1}$. In terms of components, this becomes
\begin{align}
\det(\mathcal{A}) &= \alpha_1 \alpha_2 \alpha_3\nonumber\\
        &- \alpha_1 \operatorname{Sc}(x_3 x_3^\dagger)
        - \alpha_2 \operatorname{Sc}(x_2 x_2^\dagger)
        - \alpha_3 \operatorname{Sc}(x_1 x_1^\dagger)\nonumber\\
        &+ 2\,\Re\big( x_2(x_1 x_3) \big).
\end{align}

In the complexified algebra, the determinant takes values in $\mathbb{C}$. These invariants satisfy the cubic polynomial identity~\cite{Dray:1999xe, Springer2000}
\begin{equation}
    \mathcal{A}^3-(\operatorname{tr}\mathcal{A})\mathcal{A}^2+\sigma(\mathcal{A})A-\det(\mathcal{A})\mathbf{1}=0,
    \label{eq:cubic_identity_app2}
\end{equation}
which holds equally in the complexified algebra. The corresponding characteristic equation
\begin{equation}
    \lambda^3-(\operatorname{tr}\mathcal{A})\lambda^2+\sigma(\mathcal{A})\lambda-\det(\mathcal{A})=0
\end{equation}
defines the three Jordan eigenvalues of $\mathcal{A}$. In real algebra, these are real; in the complexified algebra, they are generally complex.

The automorphism group of the real Albert algebra is
\begin{equation}
    \mathrm{Aut}(J_3(\mathbb{O})) \cong F_4,
\end{equation}
the compact exceptional Lie group of dimension 52~\cite{Baez:2001dm, Springer2000}. Automorphisms preserve the Jordan product and hence all polynomial invariants. The reduced structure group preserving the cubic norm up to scale is
\begin{equation}
    \mathrm{Str}_0(J_3(\mathbb{O})) \cong E_6,
\end{equation}
acting linearly on $J_3(\mathbb{O})$ and preserving $\det(\mathcal{A})$~\cite{Springer2000,Jacobson}. After complexification, one obtains the complex exceptional groups
\begin{equation}
    \mathrm{Aut}(J_3(\mathbb{O}_{\mathbb{C}})) \cong F_4^{\mathbb{C}},
    \qquad
    \mathrm{Str}_0(J_3(\mathbb{O}_{\mathbb{C}})) \cong E_6^{\mathbb{C}},
\end{equation}
so that the cubic spectral structure employed in the main text is naturally associated with the exceptional Lie groups $F_4$ and $E_6$.

\section{The Symmetric Cubic Representation \texorpdfstring{$\mathrm{Sym}^3(\mathbf{3})$ of $\mathrm{SU}(3)$}{Sym3(3) of SU(3)}}
\label{app:Sym3}
Let $V \cong \mathbb{C}^n$ denote the fundamental representation space of $\mathrm{SU}(n)$ with basis $\{a_1,a_2,\dots,a_n\}$. The group $\mathrm{SU}(n)$ acts on $V$ by matrix multiplication,
\begin{equation}
    v \mapsto Uv, \qquad U\in \mathrm{SU}(n), \quad v\in V .
\end{equation}
Now, consider the $m$-fold tensor product:
\begin{equation}
    V^{\otimes m} = \underbrace{V \otimes V \otimes \cdots \otimes V}_{m\ \text{times}} ,
\end{equation}
with the induced diagonal action:
\begin{equation}
    v_1\otimes v_2\otimes\cdots\otimes v_m \mapsto (Uv_1)\otimes(Uv_2)\otimes\cdots\otimes(Uv_m).
\end{equation}
The symmetric power $\mathrm{Sym}^m(V)$ is defined as the subspace of $V^{\otimes m}$ invariant under permutations of the tensor factors. Denoting by $S_m$ the symmetric group on $m$ elements, this space is
\begin{equation}
    \mathrm{Sym}^m(V) = \{\, w\in V^{\otimes m} \mid \sigma\cdot w = w,\; \forall \sigma\in S_m \,\}.
\end{equation}
Equivalently, \(\mathrm{Sym}^m(V)\) is the image of the symmetrization operator
\begin{equation}
    \mathrm{Sym} : V^{\otimes m} \to \mathrm{Sym}^m(V), \quad \mathrm{Sym}(T) = \frac{1}{m!} \sum_{\sigma \in S_m} \sigma \cdot T.
\end{equation}

A \textit{natural basis} of \(\mathrm{Sym}^m(V)\) is given by symmetrized monomials:
\begin{equation}
    a_{i_1}a_{i_2}\cdots a_{i_m} \equiv \mathrm{Sym}(a_{i_1}\otimes a_{i_2}\otimes\cdots\otimes a_{i_m}), \quad 1 \leq i_j \leq n,
\end{equation}
where the ordering of indices is irrelevant due to symmetry. The dimension of this space equals the number of monomials of degree $m$ in $n$ variables,
\begin{equation}
    \dim \mathrm{Sym}^m(V)=\binom{n+m-1}{m}.
\end{equation}
The space $\mathrm{Sym}^m(V)$ forms an irreducible representation of $\mathrm{SU}(n)$ with highest weight $m\omega_1$, where $\omega_1$ is the first fundamental weight. For the case relevant to the present work, we consider $n=3$ and $m=3$. The representation
\begin{equation}
    \mathrm{Sym}^3(\mathbb{C}^3)
\end{equation}
has dimension
\begin{equation}
    \dim \mathrm{Sym}^3(\mathbb{C}^3)=\binom{5}{3}=10 .
\end{equation}
\begin{figure}[htbp]
\centering
\includegraphics[width=0.75\columnwidth]{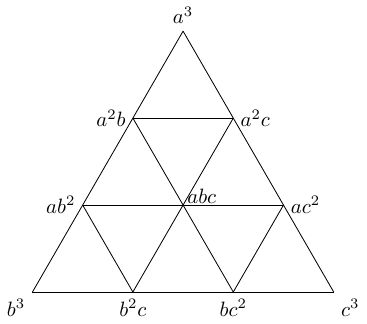}
\caption{Weight diagram of the symmetric cubic representation $\mathrm{Sym}^3(\mathbf{3})$.}
\label{fig:sym3}
\end{figure}
Choosing the basis $\{a,b,c\}$ for $\mathbb{C}^3$, a convenient basis for the cubic symmetric space is
\begin{align}
a^3,\quad b^3,\quad c^3,\quad
a^2b,\quad a^2c,\quad
b^2a,\quad b^2c,\quad
c^2a,\quad c^2b,\quad
abc.
\end{align}
This representation contains ten homogeneous cubic monomials, as illustrated in Fig.~\ref{fig:sym3}. Under the action of $U\in \mathrm{SU}(3)$ these basis elements transform according to
\begin{equation}
    a_{i_1}a_{i_2}a_{i_3}
\mapsto
    (Ua_{i_1})(Ua_{i_2})(Ua_{i_3}),
\end{equation}
which defines the symmetric cubic representation of $\mathrm{SU}(3)$.

\section{Conditional Mass-Ratio Relations in \texorpdfstring{Ref.~\cite{Singh:2025xxv}}{Ref.~[Singh:2025xxv]}}
\label{app:tp}
For completeness, we collect here the mass-ratio relations given in Ref.~\cite{Singh:2025xxv} under its specified spectral normalization, weight assignments, and cross-sector prescription. These formulae are conditional on those assumptions and are not presented here as a first-principles derivation of the physical fermion masses.
\begin{equation}
    \boxed{\delta = \sqrt{\frac{3}{8}}}
\end{equation}
\noindent\textbf{Down quarks:}
\begin{equation}
    \sqrt{\frac{m_b}{m_s}} = \frac{1+\delta}{1-\delta}(1+\delta),
    \qquad
    \sqrt{\frac{m_s}{m_d}} = \frac{1+\delta}{1-\delta}
\end{equation}
\noindent\textbf{Up quarks:}
\begin{equation}
    \sqrt{\frac{m_t}{m_c}} = \frac{\tfrac{2}{3}}{\tfrac{2}{3}-\delta},
    \qquad
    \sqrt{\frac{m_c}{m_u}} = \frac{\tfrac{2}{3}+\delta}{\tfrac{2}{3}-\delta}
\end{equation}
\noindent\textbf{Charged leptons:}
\begin{equation}
    \sqrt{\frac{m_\tau}{m_\mu}} = \frac{1+\delta}{1-\delta},
    \qquad
    \sqrt{\frac{m_\mu}{m_e}} = \frac{1+\delta}{1-\delta}\,\frac{\tfrac{1}{3}+\delta}{\delta-\tfrac{1}{3}}
\end{equation}
These expressions may be directly compared with the relations derived in Sec.~\ref{sec:spectral_construction}, where the effects of projection, normalization freedom, and octonionic phase structure modify the corresponding mass hierarchies.

\section{Exploratory: Geometric Interpretation of Square-Root Hierarchies}
\label{app:geometric}
This appendix is exploratory and presents a possible geometric interpretation of the spectral hierarchy relations derived in the main text. In this interpretation, the quantities $\sqrt{m_i}$ behave as effective geometric amplitudes associated with cubic spectral embeddings in the exceptional Jordan algebra. In this picture, fermion mass ratios arise not directly from linear spectral scales, but from quadratic combinations of underlying amplitudes determined by the exceptional Jordan spectrum.

In the cubic construction of Sec.~\ref{sec:spectral_construction}, the monomials of $\mathrm{Sym}^3(\mathbf3)$ play the role of composite spectral amplitudes built from the ordered Jordan eigenvalues $(a,b,c)$. The hierarchy relations are therefore naturally expressed at the level of amplitude ratios rather than directly in terms of masses. Physical fermion masses arise only after quadratic composition of these amplitudes, while the quantities $\sqrt{m_i}$ emerge as the corresponding linear spectral variables associated with the underlying Jordan geometry.

The cubic ladders of $\mathrm{Sym}^3(\mathbf3)$ may therefore be interpreted as defining ordered flows on the weight lattice generated by the spectral scales $(a,b,c)$. Successive hierarchy ratios correspond to additive logarithmic contrasts between neighboring amplitudes, while the prescribed weight-diagram reflection relating the selected down-quark and charged-lepton ladders acts geometrically as a discrete symmetry of the underlying spectral lattice.

Within this interpretation, each fermion sector is associated with a Hermitian Jordan element
\begin{equation}
\mathcal A_f \in J_3(\mathbb O_{\mathbb C}),
\end{equation}
whose ordered eigenvalues determine the corresponding hierarchy structure. The cubic ladder relations derived in Sec.~\ref{sec:spectral_construction} are naturally expressed in terms of square-root mass ratios, suggesting that the spectral amplitudes associated with the cubic monomials constitute the more fundamental hierarchy variables, while the physical masses arise only as secondary quadratic observables.

A possible interpretation is that the Jordan spectrum does not directly define the effective low-energy Yukawa operator, but instead determines an underlying bridge operator whose singular values correspond to the spectral amplitudes appearing in the hierarchy relations. This motivates considering a doubly exceptional structure
\begin{equation}
E_6^{(L)} \times E_6^{(R)},
\end{equation}
consisting of left and right exceptional sectors connected through an inter-sector coupling matrix. Let
\begin{equation}
Z_f:\mathcal H_R \to \mathcal H_L
\end{equation}
denote such a bridge operator between left and right fermionic sectors. The singular values of $Z_f$ are then identified with the spectral amplitudes governing the hierarchy construction. A finite Dirac operator may schematically be written in block form as
\begin{equation}
D_F=
\begin{pmatrix}
0 & Z_f \\
Z_f^\dagger & R_f
\end{pmatrix},
\end{equation}
where $R_f$ represents a heavy right-sector operator. Assuming a hierarchical regime
\begin{equation}
\|R_f\| \gg \|Z_f\|,
\end{equation}
the low-energy effective operator obtained through Schur-complement reduction takes the approximate form
\begin{equation}
Y_f^{\mathrm{eff}}
\simeq
-\, Z_f R_f^{-1} Z_f^\dagger.
\end{equation}
The effective Yukawa spectrum is therefore quadratic in the bridge amplitudes. If the singular values of $Z_f$ are
\begin{equation}
\{S_1,S_2,S_3\},
\end{equation}
then the effective fermion masses scale schematically as
\begin{equation}
m_i \sim S_i^2.
\end{equation}
Within this interpretation, the square-root hierarchy relations derived in Sec.~\ref{sec:spectral_construction} arise naturally because the Jordan spectrum controls the singular values of the bridge operator rather than the physical Yukawa eigenvalues directly. The quantity
\begin{equation}
\mathcal K_f=\frac{\sqrt{m_1m_3}}{m_2}
\end{equation}
may then be interpreted as a measure of deviation from an exact geometric hierarchy. The special value
\begin{equation}
\mathcal K_f=1
\end{equation}
corresponds to a perfectly geometric spectral ladder, while deviations from unity quantify departures from exact scaling behavior.

The discussion above should be regarded only as a possible geometric interpretation of the spectral construction developed in this work. The present paper does not provide a dynamical derivation of the doubled exceptional structure, nor of the bridge operator $Z_f$. Rather, the purpose of this section is to indicate a possible mechanism through which the square-root hierarchy structure could emerge from a more fundamental exceptional geometry. For background on exceptional geometry and Jordan algebras in particle physics, see Refs.~\cite{Baez:2001dm, Gunaydin:1973rs}. For related ideas involving doubled structures and effective quadratic reductions, see Ref.~\cite{Chamseddine:2022rnn}.

\section{Spin(9)-Adapted Peirce Decomposition of the Cubic Jordan Invariant}
\label{app:spin9_peirce}
The one-generation sector reviewed in Sec.~\ref{sec:fermion_generations} admits a Clifford-algebraic description through the ladder operators $\{\alpha_i,\alpha_i^\dagger\}_{i=1}^3$, which generate
\begin{equation}
\mathbb C\ell(6)\cong M_8(\mathbb C).
\end{equation}
One generation of fermions is organized as a minimal left ideal
\begin{equation}
\mathbb C\ell(6)\cdot \pi,
\end{equation}
with the number operator $N$ determining the charge grading $Q=N/3$. The three-generation hierarchy construction, however, is governed not by the Clifford algebra itself, but by the exceptional Jordan algebra
\begin{equation}
J_3(\mathbb O_{\mathbb C}),
\qquad
\mathbb O_{\mathbb C}=\mathbb C\otimes\mathbb O,
\end{equation}
which plays a central role in exceptional Jordan geometry and octonionic algebraic constructions~\cite{Baez:2001dm, Springer2000}. The Albert algebra is exceptional and cannot be globally reduced to a Clifford-algebraic quadratic structure~\cite{Springer2000}. Nevertheless, after choosing a Peirce decomposition adapted to a primitive idempotent, a natural Spin(9) structure emerges inside the cubic Jordan invariant.

\subsection{Spin(9)-Adapted Decomposition}
Choose the primitive idempotent
\begin{equation}
e_{11}=
\begin{pmatrix}
1&0&0\\
0&0&0\\
0&0&0
\end{pmatrix}.
\end{equation}
A general element of \(J_3(\mathbb O_{\mathbb C})\) may then be written as
\begin{equation}
\mathcal A=
\begin{pmatrix}
\xi & x & y^\dagger\\
x^\dagger & \alpha & z\\
y & z^\dagger & \beta
\end{pmatrix},
\qquad
\xi,\alpha,\beta\in\mathbb C,
\qquad
x,y,z\in\mathbb O_{\mathbb C}.
\end{equation}
Equivalently,
\begin{equation}
\mathcal A=
\begin{pmatrix}
\xi & \phi^\dagger\\
\phi & \mathcal{V}
\end{pmatrix},
\qquad
\phi=
\begin{pmatrix}
x\\ y
\end{pmatrix},
\qquad
\mathcal{V}=
\begin{pmatrix}
\alpha & z\\
z^\dagger & \beta
\end{pmatrix}
\in J_2(\mathbb O_{\mathbb C}).
\end{equation}
The corresponding Peirce decomposition is
\begin{equation}
J_3(\mathbb O_{\mathbb C}) \cong \mathbb C e_{11} \oplus J_2(\mathbb O_{\mathbb C}) \oplus \mathbb O_{\mathbb C}^{\,2}.
\end{equation}
Under the Spin(9) subgroup stabilizing \(e_{11}\)~\cite{Springer2000, Bengtsson:1987si},
\begin{equation}
J_2(\mathbb O_{\mathbb C}) \cong \mathbf 1\oplus \mathbf 9,
\end{equation}
while \(\mathbb O_{\mathbb C}^{\,2}\) carries the complexified \(\mathbf{16}\) spinor representation. Hence,
\begin{equation}
\mathbf{27} = \mathbf 1 \oplus (\mathbf 1\oplus\mathbf 9) \oplus \mathbf{16}.
\end{equation}
Removing the trace direction gives the traceless decomposition
\begin{equation}
\mathbf{26} = \mathbf 1\oplus\mathbf 9\oplus\mathbf{16}.
\end{equation}
Writing
\begin{equation}
\mathcal V=t\,\mathbf 1_2+ \mathcal W,
\qquad
t=\frac{\alpha+\beta}{2},
\end{equation}
with
\begin{equation}
\mathcal W=
\begin{pmatrix}
u & z\\
z^\dagger & -u
\end{pmatrix},
\qquad
u=\frac{\alpha-\beta}{2},
\end{equation}
the scalar \(t\) represents the lower-block trace component, while $\mathcal{W}$ transforms in the vector representation appearing in the Spin(9)-adapted decomposition.

\subsection{Spin(9) Structure of the Cubic Jordan Norm}
For the Hermitian element \(\mathcal A\), the cubic Jordan norm~\cite{Dray:1999xe} is
\begin{equation}
\det \mathcal A = \xi\alpha\beta -\xi\,\operatorname{Sc}(zz^\dagger) -\alpha\,\operatorname{Sc}(yy^\dagger) -\beta\,\operatorname{Sc}(xx^\dagger) +2\,\Re\!\left(y(xz)\right).
\end{equation}
Equivalently,
\begin{equation}
\det \mathcal A = \xi\,\det \mathcal V - \mathcal Q_9(\phi,\mathcal V,\phi),
\end{equation}
where
\begin{equation}
\det \mathcal V = \alpha\beta-\operatorname{Sc}(zz^\dagger),
\end{equation}
and
\begin{equation}
\mathcal Q_9(\phi,\mathcal V,\phi) = \beta\,\operatorname{Sc}(xx^\dagger) + \alpha\,\operatorname{Sc}(yy^\dagger) - 2\,\Re\!\left(y(xz)\right).
\end{equation}
Using the decomposition \(\mathcal V=t\mathbf 1_2+\mathcal W\),
\begin{equation}
\begin{aligned}
\mathcal Q_9(\phi,\mathcal V,\phi)
&=
t\Big[
\operatorname{Sc}(xx^\dagger)
+
\operatorname{Sc}(yy^\dagger)
\Big] \\
&\quad
+
u\Big[
\operatorname{Sc}(yy^\dagger)
-
\operatorname{Sc}(xx^\dagger)
\Big]
-
2\,\Re\!\left(y(xz)\right).
\end{aligned}
\end{equation}
The first term couples the scalar component to the spinor norm, while the remaining terms encode the nontrivial trilinear structure of the cubic Jordan invariant. After choosing a preferred idempotent and the associated Spin(9)-adapted decomposition, the octonionic triple product
\begin{equation}
\tau=\Re\!\left(y(xz)\right)
\end{equation}
admits a local interpretation analogous to a Spin(8)-type spinor-vector-spinor coupling inherited from triality. In this polarized decomposition, the \(z\)-dependent contribution may be viewed as playing the role of the vector component, although this identification is not globally invariant under the full exceptional symmetry.

\subsection{Symmetric Sector and Canonical Spread}
For the symmetric Jordan element used in the main text,
\begin{equation}
\mathcal A_f=
\begin{pmatrix}
rs_f & x_f & y_f^\dagger\\
x_f^\dagger & rs_f & z_f\\
y_f & z_f^\dagger & rs_f
\end{pmatrix},
\end{equation}
the determinant reduces to
\begin{equation}
\det \mathcal A_f=r^3s_f^3-rs_f\delta_f^2+2\tau_f,
\end{equation}
with
\begin{equation}
\delta_f^2
=
\operatorname{Sc}(x_fx_f^\dagger)
+
\operatorname{Sc}(y_fy_f^\dagger)
+
\operatorname{Sc}(z_fz_f^\dagger),
\end{equation}
and
\begin{equation}
\tau_f = \Re\!\left(y_f(x_fz_f)\right).
\end{equation}
This reproduces the invariant structure used in the main text and provides a Spin(9)-adapted interpretation of \(\tau_f\) as the cubic octonionic contribution controlling spectral asymmetry. In the symmetric normalization,
\begin{equation}
\operatorname{Sc}(x_fx_f^\dagger)
=
\operatorname{Sc}(y_fy_f^\dagger)
=
\operatorname{Sc}(z_fz_f^\dagger)
=
\frac18,
\end{equation}
so that
\begin{equation}
\delta_f^2=\frac38.
\end{equation}
Within the Peirce decomposition, this canonical spread combines both the spinor contribution
\begin{equation}
\operatorname{Sc}(x_fx_f^\dagger)
+
\operatorname{Sc}(y_fy_f^\dagger)
\end{equation}
and the vector contribution
\begin{equation}
\operatorname{Sc}(z_fz_f^\dagger).
\end{equation}
Thus \(\delta_f^2=3/8\) measures the normalized quadratic size of the off-diagonal Jordan data within the symmetric sector.

\subsection{Interpretation and Limitations}
The Spin(9)-adapted decomposition clarifies the algebraic role of the octonionic triple product in the spectral construction: the invariant
\begin{equation}
\tau_f=\Re\!\left(y_f(x_fz_f)\right)
\end{equation}
is the component of the cubic Jordan norm associated with spectral asymmetry away from the reflection-symmetric case \(\tau_f=0\). However, this decomposition does not by itself determine the effective phase parameters appearing in the phenomenological fit. In particular, the parametrization
\begin{equation}
\tau_e=\frac{1}{64}\cos\Phi_e
\end{equation}
should be regarded as an effective spectral description of the relevant octonionic triple product rather than as a quantized prediction of the Spin(9) structure alone. Additional dynamical or representation-theoretic input would be required to derive discrete phase values. No such additional structure is assumed in the present framework.

\bibliography{references}

\end{document}